\documentclass[twocolumn,showpacs,preprintnumbers,amsmath,amssymb]{revtex4}

\usepackage{graphicx}
\usepackage{dcolumn}
\usepackage{bm}
\usepackage{grffile}
\begin{document}

\title{What densities can be effectively probed in quasifree
  single-nucleon knockout reactions?}

\author{Jan Ryckebusch}
\email{Jan.Ryckebusch@UGent.be}
\author{Wim Cosyn}
\email{Wim.Cosyn@UGent.be}
\author{Maarten Vanhalst}
\email{Maarten.Vanhalst@UGent.be}

\affiliation{Department of Physics and Astronomy,\\
 Ghent University, Proeftuinstraat 86, B-9000 Gent, Belgium}
\date{\today}

\begin{abstract}
We address the issue whether quasifree single-nucleon knockout
measurements carry sufficient information about the nuclear interior.
To this end, we present comparisons of the reaction probability
densities for $A(e,e'p)$ and $A(p,2p)$ in quasifree
kinematics for the target nuclei $^{4}$He, $^{12}$C, $^{56}$Fe, and
$^{208}$Pb.  We adopt a comprehensive framework based on the impulse
approximation and on a relativized extension of Glauber
multiple-scattering reaction theory in which the medium effects
related to short-range correlations (SRC) are implemented. It is
demonstrated that SRC weaken the effect of attenuation. For light
target nuclei, both the quasifree $(p,2p)$ and $(e,e'p)$ can probe
average densities of the same order as nuclear saturation density $\rho
_{0}$. For heavy nuclei like $^{208}$Pb, the probed average densities are
smaller than $0.1\rho _{0}$ and the $(e,e'p)$ reaction is far more
efficient in probing the bulk regions than $(p,2p)$. 

\end{abstract}

%


\maketitle 

\section{Introduction}

Single-nucleon knockout reactions from nuclei in quasifree kinematics
continue to be major source of information of the mean-field
properties of nuclei.  In a quasifree $A(e,e'p)$ reaction, a bound
nucleon in the target nucleus $A$ is subjected to a electron-nucleon
interaction and ejected, thereby leaving the residual nucleus in a
low-lying hole state. The shape of the measured differential cross
sections and the knowledge about the energy of the residual nucleus, 
allows one to determine the quantum numbers of the nucleon that was
struck by the virtual photon. Electroinduced single-proton knockout
experiments with stable nuclear targets have systematically pointed
towards the validity but also the limitations of the mean-field
picture for understanding nuclei
\cite{ISI:A1996BF20A00002,Pandharipande:1997zz}.

The measured quasifree $A(e,e'p)$ differential cross sections provided
evidence for the mean-field picture in that the extracted momentum
distributions for the bound nucleons could be modeled with mean-field
single-particle wave functions. It should be stressed that the
measured momentum distributions cannot be directly related to the
single-particle wave functions in momentum space, but are distorted in
the sense that they are affected by the attenuation effects of the
nuclear medium on the ejected proton.  The systematic observation that the
extracted spectroscopic factors from $A(e,e'p)$ studies are
substantially smaller than the predicted single-particle level
occupancies provide evidence for the limitations of
the mean-field picture 
\cite{Lapikas:1003zz}. The extracted spectroscopic factors are defined
as the overall normalization factor between the measured and the
computed differential  cross sections. The computed $A(e,e'p)$
observables are based on models that make assumptions for the
electron-nucleus interaction, the nuclear wave functions, and the
effect of nuclear attenuation on the ejected proton. The latter effect
is often referred to as final-state interactions (FSI). The
credibility of a reaction model for $A(e,e'p)$ depends on its ability
to describe for example the differential cross sections and polarization
observables for a number of
target nuclei. An interesting question is whether the extracted
normalization factors can be related to the shell occupancy in a
model-independent fashion \cite{Furnstahl:2001xq,Mukhamedzhanov:2010hc}.

An alternative method to gain access to the mean-field properties of
nuclei is the $A(p,2p)$ reaction \cite{RevModPhys.45.6}.  For the
study of stable nuclei one may prefer the $A(e,e'p)$ reaction as it
leads to some reduced parameter dependence of the extracted results,
like spectroscopic factors. In the first place, this is due to the
electromagnetic character of the interaction vertex. Second, for
obvious reasons the sensitivity to the modeled nuclear attenuation is
larger for $(p,2p)$ than for $(e,e'p)$.  Electron scattering
experiments from unstable nuclei are a real technical challenge and 
could for example be performed at an $eA$ collider like the one which
is on the drawing table for the ELISe (ELectron-Ion Scattering in a
Storage Ring) project at FAIR \cite{Elise}.  In inverse kinematics
(i.e. the $p(A,2p)A-1$ process) the $A(p,2p)$ reaction offers great
perspectives to investigate the mean-field properties of unstable
nuclei \cite{Chulkov200543}. One of the fundamental questions, for
example, we need to ask ourselves is how the single-particle
properties of nuclei evolve as a function of the proton-to-neutron
ratio. Recent studies \cite{Gade:2008} with the heavy-ion-induced
nucleon-knockout reactions $^{9}$Be$(A,A-1)X$, suggest that the
deduced spectroscopic factors for single-proton and single-neutron
knockout are dramatically dependent on the asymmetry of the proton and
neutron Fermi surface. The $^{9}$Be$(A,A-1)X$ process is extremely surface
dominated and the above-mentioned results with regard to spectroscopic
factors await confirmation with a reaction probe which is more
efficient in probing the nuclear interior.

\begin{figure}
 \centering
 \includegraphics[viewport=11 41 543 263,clip,width=0.45\textwidth]{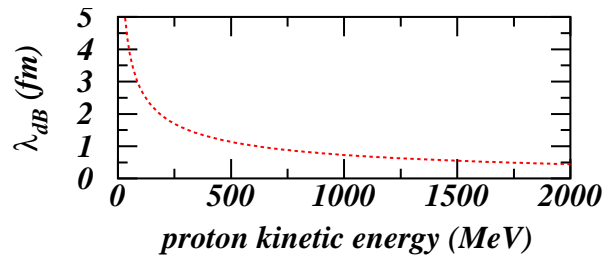}
 \caption{The de Broglie wavelength ($\lambda _{dB}$) of a proton as a
 function of its kinetic energy.}
 \label{fig:wavelen}
\end{figure}

The scientific potential of the $(p,2p)$ reaction in inverse
kinematics to study the mass number dependence of mean-field
properties, is nicely illustrated in Ref.~\cite{Kobayashi2008}. There,
results of $(p,2p)$ measurements on the eight carbon isotopes
$^{9-16}$C are presented.  The measurements provide empirical
information about the mass-number dependence of the weakly-bound and
inner-shell protons. It is shown that the systematics of the momentum
distributions, separation energies, and spectroscopic factors as a
function of the mass number can be studied. The $(p,2p)$ measurements
of Ref.~\cite{Kobayashi2008} are performed at energies 
of 250~MeV/A. In the foreseeable future, quasifree $(p,2p)$
measurements for protons of several hundreds of MeV/A will become
feasible with high-energy heavy ion beams at the accelerator complex
FAIR \cite{Aumann2007}. The conditions of high energy are beneficial
from the theoretical point of view. First, under high-energy
conditions the $(p,2p)$ interaction range is small compared to the
size of the target nucleus and one can make use of the zero-range
approximation when modeling the proton-nucleus interaction
vertex. Second, for fast continuum nucleons with a sufficiently small
de Broglie wavelength, the effect of attenuation can be appropriately
and accurately computed in the semi-classical eikonal approximation
\cite{Benhar:1996dd} even when adopting a Dirac treatment of the
nucleons in the initial and final channel \cite{Amado:1984dy}. From
Fig.~\ref{fig:wavelen} it is clear that the proton de Broglie
wavelength drops below 1 fm for nucleon kinetic energies larger than
about 500~MeV. Several studies \cite{Lava:2004zi} have pointed towards
the applicability of the eikonal method up to remarkably low nucleon
kinetic energies of about 250~MeV. 
  
Just as for the $A(e,e'p)$ reaction, the observables from 
 $A(p,2p)$ measurements are convoluted in that nuclear
attenuation does not allow one to relate the measured ejected nucleon
properties directly to the physics at the interaction point. With one
proton subject to initial-state interactions (ISI), and two protons
subject to FSI, the development of a reliable reaction theory is of
the utmost importance for a quantitative analysis of the $A(p,2p)$
data \cite{Benhar:1996dd, PhysRevC.74.064608, overmeire:064603}.  It
is well established that nuclear attenuation tends to make the
detected signals less sensitive to the high-density regions of the
target. Accordingly, one point of concern is whether or not one
may learn something about the bulk properties of nuclei from
$A(p,2p)$. Or, in other words, carry the ejected nucleons information
about the interior of the system or do they mostly originate from the
peripheral regions of the nucleus? In this paper we wish to address
this issue.  We attempt to quantify what regions of the target nucleus
can be probed in $A(p,2p)$.  For the sake of reference, we also add
results for $A(e,e'p)$. The latter serve as a benchmark, as with
$A(e,e'p)$ a lot has been learned about single-particle properties of
stable nuclei during the last couple of decades. We focus on
single-nucleon knockout reactions with nucleon kinetic energies of
several hundreds of MeV. At those energies the Glauber multiple
scattering framework is both appropriate and accurate to model nuclear
attenuation.

In Sec.~\ref{subsec:crosssection}, we outline the necessary formalism
for quasifree single-nucleon knockout in a relativistic and
cross-section factorized framework. In Sec. \ref{subsec:rmsga} a
relativized version of Glauber multiple scattering theory is
introduced and it is pointed out how it can be corrected for the
medium effects related to SRC. The results of the the numerical
calculations are discussed in Sec.~\ref{sec:results} and a summary is
given in Sec.~\ref{sec:conclusions}.

\section{Theory}
\label{sec:theory}
\subsection{Cross sections}
\label{subsec:crosssection}
The theoretical calculations for the $A(p,pN)$ and $A(e,e'N)$
reactions presented here are performed with a factorized form for the
cross sections. They represent very useful zero-th order
approximations and are often used as a starting point for the
interpretation of the
data.  The factorization is largely based on the impulse
approximation (IA) which develops out of a reaction picture in which
$(A-1)$ nucleonic degrees-of-freedom are frozen during the interaction
of the external probe and the target.  In the IA, the role of the
spectator $(A-1)$ nucleonic degrees-of-freedom is restricted to
distorting the waves of the impinging and ejected nucleons.  In
addition, the factorization allows for a more direct comparison
between the $(p,2p)$ and the $(e,e'p)$, as their differential cross
sections become proportional to a distorted momentum distribution that
is related to the probability of finding a nucleon with well-defined
quantum numbers and a certain momentum in the target nucleus.  First,
we derive a factorized expression for the $A(e,e'p)$ and the $A(p,2p)$
cross section.

Consider the $A(e,e'N)A-1$ process and define the corresponding
kinematic variables of the impinging electron, the target nucleus, the
scattered electron, the residual nucleus, and the ejected nucleon as 
\begin{eqnarray}
K ^{\mu} \left( \epsilon, \vec{k} \right) & + &  
K ^{\mu} _ {A} \left( E_{A}, \vec{k} _{A} \right)
\longrightarrow 
K ^{\mu} \left( \epsilon ', \vec{k} ' \right) 
\nonumber \\
& + & 
K ^{\mu} _ {A-1} \left( E_{A-1} , \vec{k} _{A-1} \right)  +
K ^{\mu} _ {N} \left( E_{N} , \vec{k} _{N} \right)  \; \; \; \; .
\label{eq:eepkin}
\end{eqnarray} 
The fivefold differential cross section in the lab frame 
adopts the form \cite{Ryckebusch:2003fc}
\begin{equation}
\frac { d ^{5} \sigma } { d \epsilon ' d \Omega _ {e'} d \Omega _{N} } = 
\frac { m_{e} ^{2} k ' k_{N} M_{A-1} M_{N} } 
{ \left( 2 \pi \right) ^{5} \beta _{eA} \epsilon M_{A} } \left(
f_{rec}^{eA} \right) ^{-1} 
\overline{ \sum _{if} }\left| \mathcal{M} _{fi} ^{(e,e'p)}  \right| ^{2}  \; ,
\label{eq:eepcross}
\end{equation}
where $\beta _{eA} = \frac {k} {\epsilon} \approx 1 $ is the relative
velocity and $ \overline{ \sum _{if} } $ corresponds to the
appropriate average over initial states and sum over final states.  We
assume that all wave functions entering the computation of the
reaction amplitude $ \mathcal{M} _{fi} ^{(e,e'p)} $ are normalized to
unity. Throughout this work, the adopted normalization convention for
the Dirac spinors is 
$ \bar{u } \left( \vec{k}, m_s \right) u
\left( \vec{k}, m_s \right) = 1 $.  
The recoil factor $
f_{rec}^{eA} $ in Eq.~ (\ref{eq:eepcross}) reads
\begin{equation}
f_{rec}^{eA} = \frac {E_{A-1}} {M_{A} } \left(
1 + \frac {E_{N}}  {E_{A-1} } 
\left( 1 - \frac { \vec {q} \cdot \vec{k}_{N}  } {k_{N} ^{2} }
\right) \right) \; , 
\label{eq:eeprec}
\end{equation}
where the momentum transfer has been defined as
\begin{equation}
\vec{q} = \vec{k} - \vec{k} ' \; .
\end{equation}

For the kinematics of the $A(p,2p)A-1$ we adopt the following
conventions
\begin{eqnarray}
P_{1} ^{\mu} \left( E_{p1}, \vec{p}_{1} \right) & + & 
K ^{\mu} _ {A} \left( E_{A}, \vec{k} _{A} \right)
\rightarrow 
K _{1} ^{\mu} \left( E_{k1}, \vec{k} _{1} \right) 
\nonumber \\
& + & 
K _{2} ^{\mu} \left( E_{k2}, \vec{k} _{2} \right) + 
K ^{\mu} _ {A-1} \left( E_{A-1} , \vec{k} _{A-1} \right) 
  \; , \; \; \; \;
\label{eq:pAkin}
\end{eqnarray} 
where $ P_{1} ^{\mu} $ refers to the impinging nucleon, and $\left( K _{1}
^{\mu}, K _{2} ^{\mu} \right) $ to the pair of ejected nucleons. The
fivefold differential equation for the $A(p,2p)A-1$ process can be
straightforwardly derived from
the $A(e,e'p)$  one (Eq.~(\ref{eq:eepcross})) by means of the
following substitutions ($m_{e} \rightarrow M_{N}, \epsilon \rightarrow
E_{p1}, k ' \rightarrow k_{1}, k_{N} \rightarrow k_{2} $) and this
results in
\begin{equation}
\frac { d ^{5} \sigma } { d E_{k1} d \Omega _ {1} d \Omega _{2} } = 
\frac { M_{N} ^{3} k _{1} k_{2} M_{A-1} } 
{ \left( 2 \pi \right) ^{5} \beta _{pA} E_{p1} M_{A} } 
\left( f_{rec}^{pA} \right) ^{-1} 
\overline{ \sum _{if} }\left| \mathcal{M} _{fi} ^{(p,2p) }  \right| ^{2}  \; ,
\label{eq:pAcross}
\end{equation}
where the recoil factor reads
\begin{equation}
f_{rec}^{pA} = \frac {E_{A-1}} {M_{A} } \left(
1 + \frac {E_{k2}}  {E_{A-1} } 
\left( 1 - \frac { \vec {q} \cdot \vec{k_{2} } } {k_{2} ^{2} }
\right) \right) \; .
\label{eq:pArec}
\end{equation}
In this expression the momentum transfer is defined as
\begin{equation}
\vec{q} = \vec{p}_{1} - \vec{k} _{1} \; .
\end{equation}
In the laboratory frame one obtains the following expression for the
relative velocity $\beta _{pA} = \frac {p_{1}} {E_{p1}} $.

The squared amplitude $ \overline{ \sum _{if} } \left| \mathcal{M}
_{fi} ^{(p,2p) } \right| ^{2} $ can be related to the free
proton-proton cross section after making several assumptions. They are
pointed out in great detail in Ref.~\cite{overmeire:064603} and
include the neglect of the negative-energy components in the
proton-proton scattering amplitude. Further, one assumes that the
operator describing the collision of the impinging proton
$\vec{p}_{1}$ and the bound proton $\vec{p}_{m}$ resulting can be
described in terms of the on-shell proton-proton scattering amplitude.
Now, we sketch how one can arrive at a factorized expression for the
amplitude $ \mathcal{M} _{fi} ^{(p,2p) } $. We denote the position
coordinates of the impinging proton as $\vec{r}_{0}$, of the ejected
protons as $\left( \vec{r}_{0} , \vec{r}_{1} \right) $, and of the
nucleons in the target nucleus as ($\vec{r}_{1}, \ldots ,
\vec{r}_{A}$).  In the eikonal approach, the attenuation can be
accounted for by means of a multiplicative factor to be applied to a
plane-wave wave function.  Accordingly, the relativistic distorted wave
function of the impinging proton reads
\begin{eqnarray}
\phi _{\vec{p} _ 1, m_{s1} } ^{D} 
& = & 
\widehat{S} _{p1} \left( \vec{r}_{0} , \vec{r}_{2}, \ldots ,
\vec{r}_{A} \right) 
\sqrt { \frac { E + M} {2 M} } 
\left[
\begin{array}{c} 
1 \\
\frac { \vec{\sigma} \cdot \vec{p} _{1} } { E + M} 
\end{array}
\right] \nonumber \\
& & \times e ^{ i \vec{p} _{1} \cdot \vec{r} _ {0} } \chi _ {\frac {1} {2} m_{s1} } 
= \widehat{S} _{p1}  e ^{ i \vec{p} _{1} \cdot \vec{r} _ {0} } 
u \left( \vec{p}_{1} , m_{s1} \right) \; ,
\label{eq:wavimpin}
\end{eqnarray} 
where $u(\vec{p},s)$ is a four-component free-particle Dirac spinor.
For the distorted wave
functions $\phi _{\vec{k} _ {1}, m'_{s1}} ^{D} $ and $\phi _{\vec{k} _
  {2}, m'_{s2}} ^{D} $ of the ejected protons, similar
expressions hold.  
In the above expression, the $A$-body operator $\widehat{S} _{p1}$ is
fully responsible for the effect of attenuation. In this work, we
compute the effect of attenuation in a relativized Glauber model. More
details will be provided in the next subsection \ref{subsec:rmsga}.
 The $\widehat{S} _{p1}$ is a
two-by-two matrix which acts on the Pauli spinors. Here, we assume
that $\widehat{S} _{p1}$ is a diagonal matrix which amounts to
neglecting the spin-dependent attenuation mechanisms.  Indeed, the
central component of the nucleon-nucleon scattering amplitude accounts
for the major impact of nuclear attenuation in high-energy proton-nucleus
collisions \cite{Alkhazov:1978et}. Recent investigations
\cite{Jeschonnek:2008zg} have clarified the role of the spin-dependent
terms in the nucleon-nucleon scattering amplitude for the attenuation
effects in $D(e,e'p)n$. The central component was found to dominate
the attenuation for most observables, and in particular for the
differential cross sections at low missing momenta. Accordingly, we
deem that inclusion of the central component is sufficiently accurate
for our current purposes, namely making a comparative study of the
density dependence of the $(e,e'p)$ and $ (p, 2p) $ reaction
throughout the mass table.

We now proceed with the derivation of a factorized expression for the
$(e,e'p)$ and $(p,2p) $ cross section.  In what follows we describe
the wave function of the target nucleus by a normalized Slater
determinant $\left| \alpha _{1} \alpha _{2} \ldots \alpha _{A} \right>
$ where the $\alpha _{i}$ refer to the quantum numbers of the occupied
single-particle states.  With the distorted wave functions of
Eq.~(\ref{eq:wavimpin}) we get after neglecting the negative-energy
projection term the following expression \cite{overmeire:064603}
\begin{eqnarray}
& & \mathcal{M}_{fi} ^{(p,2p) }   \approx    
\sum _{s}  \int d \vec{r} 
\widehat{\mathcal{S}}_{\text{RMSGA}}^ {(p,2p)} (\vec{r})
e ^{ - i \vec{p}_{m} \cdot \vec{r} }
\overline{u} \left( \vec{p}_{m} , s \right)
\phi_{ n \kappa m } (\vec{r}) 
\nonumber \\
& & \times
\biggl[
u ^{\dagger} \left( \vec{k}_{1} , m '_{s1} \right) 
u ^{\dagger} \left( \vec{k}_{2} , m '_{s2} \right)
\widehat{F}_{pp}  
u  \left( \vec{p}_{1} , m _{s1} \right) 
u  \left( \vec{p}_{m} , s  \right)
\biggr] \; \; \; \; ,
\nonumber \\
& &
\label{eq:ampp2pfac}
\end{eqnarray}
where $\widehat{F} _{pp} $ is the $pp$ scattering amplitude in momentum
space. The quantum numbers $(n \kappa m) = \alpha_{1}$
determine the orbit of the struck nucleon which is described by a
relativistic single-particle wave function $ \phi_{ n \kappa m }
(\vec{r}) $.  The missing momentum
$\vec{p}_{m}$ is determined by the difference between the
asymptotic three-momentum of the ejected nucleon $\vec{k}_{2}$ and the
three-momentum transfer $ \vec{q} = \vec{p}_1 - \vec{k}_1$
\begin{equation}
\vec{p}_{m} = \vec{k}_{2} + \vec{k}_1 - \vec{p}_1 = - \vec{k} _{A-1}
\; . 
\end{equation} 
In the absence of nuclear attenuation, the missing momentum equals 
the momentum of the bound nucleon with quantum numbers $(n \kappa m)$ 
which collides with the proton beam.   
In the above equation (\ref{eq:ampp2pfac}) we have introduced an operator
which accounts for the ISI/FSI
\cite{overmeire:064603}
\begin{eqnarray}
\widehat{\mathcal{S}}_{\text{RMSGA}}^ {(p,2p)} (\vec{r}) & = & 
\prod _{i=2} ^{i=A} \int d \vec{r} _{i}
\left| \phi _{\alpha _{i} } \left( \vec{r} _{i} \right) \right| ^{2} 
\widehat{S} _{p1} \left( \vec{r} , \vec{r}_{2}, \ldots ,
\vec{r}_{A} \right) 
\nonumber \\
& \times & 
\widehat{S} _{k1}   \left( \vec{r} , \vec{r}_{2}, \ldots ,
\vec{r}_{A} \right) 
\widehat{S} _{k2}   \left( \vec{r} , \vec{r}_{2}, \ldots ,
\vec{r}_{A} \right) \; \; \; \; \; ,
\label{eq:srmsgap2p}
\end{eqnarray}
a multi-dimensional
convolution over the squared wave functions of the spectator nucleons
and a product of the scalar operators $ \widehat{S}_{k} $ for the
impinging proton and two ejected protons

With the aid of Eq.~ (\ref{eq:ampp2pfac}) one arrives
at the following factorized form for the five-fold $A(p,2p)$
differential cross section of Eq.~(\ref{eq:pAcross})
\begin{eqnarray}
\frac { d ^{5} \sigma } { d E_{k1} d \Omega _ {1} d \Omega _{2} }
& \approx &  
\frac { \left( 2 \pi \right) ^{3}   s k _{1} k_{2} M_{A-1} } 
{  M_{N} p_{1} M_{A} } 
\left( f_{rec}^{pA} \right) ^{-1}
\nonumber \\
& & \times  
\left( \frac { d \sigma ^{pp} } {d \Omega } \right) _{c.m.} 
S_{n \kappa}
\rho _{ n \kappa  } ^{D} (\vec{p}_{m}) \; ,
\label{eq:pAcrossfac}
\end{eqnarray}
where $ \left( \frac { d \sigma ^{pp} } {d \Omega } \right) _{c.m.} $
is the center-of-mass cross section for $pp$ scattering at an
invariant energy $W = \sqrt{s}$.  The $ S_{n \kappa} $ is the spectroscopic
factor ( $ 0 < S_{n \kappa} \le 1 $) that is related to the occupancy of the
orbit $(n \kappa)$ in the ground state of the target nucleus.      
In the above expression the distorted momentum distribution 
$ \rho _{n \kappa } ^{D} (\vec{p}_m) $ adopts the form \cite{Cosyn:2009bi} 
\begin{eqnarray}
\rho _{ n \kappa  } ^{D} (\vec{p}_{m}) & = &
\sum_{s , m}
\left|  \int d \vec{r}
\frac { 
 e^{-i\vec{p} _{m} \cdot \vec{r}}
} 
{(2\pi)^{3/2}} 
\bar{u}(\vec{p}_m, s) 
\widehat{\mathcal{S}}_{\text{RMSGA}} ^{(p,2p)} (\vec{r})
\phi_{ n \kappa m } (\vec{r})
\right|^2 \; , 
\nonumber \\
& = & \sum _{s, m} \left( {\phi_ {n \kappa m }  ^{D} } 
\left( \vec{p} _{m} \right)  
\right) ^{\dagger} 
{ \phi_ {n \kappa m }  ^{D} \left( \vec{p} _{m} \right)  }
\; ,
\nonumber \\
& = & \frac {1} {2} \int d r \int d \theta 
\biggl[ \sum _{s,m} \biggl( \left( { D(r, \theta) } \right) ^{\dagger} {\phi_ {n 
      \kappa m  }
    ^D \left( \vec{p} _{m} \right)  } 
\nonumber \\
& & + 
 { D(r, \theta) } \left( {\phi_ { n \kappa m  } ^D \left( \vec{p} _{m} \right)   } \right)
 ^{\dagger} \biggr) \biggr] \; ,   
\nonumber \\
& \equiv & \int d r \int d \theta \delta \left( r, \theta \right) \; 
\; ,
\label{eq:rhormsga}
\end{eqnarray}
%
where 
\begin{equation}
D(r, \theta) = \int d \phi \; r^2 \; \sin \theta \frac { 
 e^{-i\vec{p} _{m} \cdot \vec{r}}
} 
{(2\pi)^{3}} 
\bar{u}(\vec{p}_m, s) 
\widehat{\mathcal{S}}_{\text{RMSGA}}^{(p,2p)} (\vec{r})
\phi_{ n \kappa m } (\vec{r}) \; .
 \end{equation}

Working along similar lines one can derive a factorized form of the $A(e,e'p)$ 
differential cross section \cite{Caballero:1997gc}
\begin{equation}
 \frac { d ^{5} \sigma } { d \epsilon ' d \Omega _ {e'} d \Omega _{N} } = 
\frac { k_{N} M_{A-1} M_{N} } 
{ M_{A} } \left(
f_{rec}^{eA} \right) ^{-1}  \sigma ^{ ep} 
S_ {n \kappa} \rho _{ n \kappa  } ^{D} (\vec{p}_{m})
 \; ,
\label{eq:factorizedeep}
\end{equation}
where $ \sigma ^{e p } $ is the off-shell electron-proton cross
section obtained from positive-energy projections.  
The distorted momentum distribution is defined as in Eq.~(\ref{eq:rhormsga}) 
whereby $ \widehat{\mathcal{S}}_{\text{RMSGA}} ^{(p,2p)}$ is replaced by 
$\widehat{\mathcal{S}}_{\text{RMSGA}} ^{(e,e'p)}$ that adopts the form
\cite{Ryckebusch:2003fc}.
\begin{equation}
\widehat{\mathcal{S}}_{\text{RMSGA}}^ {(e,e'p)} (\vec{r})  =  
\prod _{i=2} ^{i=A} \int d \vec{r} _{i}
\left| \phi _{\alpha _{i} } \left( \vec{r} _{i} \right) \right| ^{2} 
\widehat{S} _{k1}   \left( \vec{r} , \vec{r}_{2}, \ldots ,
\vec{r}_{A} \right) \; .
\label{eq:srmsgaeep}
\end{equation}

%

\subsection{Relativistic multiple-scattering Glauber approximation}
\label{subsec:rmsga}
The quantity $\delta ( r , \theta) dr d \theta $ defined in
Eq.~(\ref{eq:rhormsga}) encodes the contribution from an infinitesimal
interval $\left[ r + dr , \theta + d \theta \right] $  to 
the cross section for 
a quasifree $p + A
\rightarrow p + p + A-1 $ process that leaves the residual nucleus in
a hole state determined by the quantum numbers $( n \kappa m) $
\cite{PhysRevLett.78.1014}.  The eikonal operator $
\widehat{\mathcal{S}}_{\text{RMSGA}}^ {(p,2p)}(\vec{r}) $ of
Eq.~(\ref{eq:srmsgap2p}) receives contributions from the impinging
($\widehat{S} _{p1}$) and the two ejected ($\widehat{S} _{k1}, \widehat{S}
_{k2}) $ protons.  In Refs.~\cite{Ryckebusch:2003fc} and
\cite{VanOvermeire:2007zy} we developed a relativized version of
Glauber multiple-scattering theory. In this so-called RMSGA approach
the eikonal phases are diagonal $2 \times 2$ matrices
\begin{eqnarray}
\widehat{S} _{p1} \left( \vec{r} \left( \vec{b},z \right) 
, \vec{r} _{2} , \ldots \vec{r} _{A}
\right) & = & \prod _{j=2} ^{j=A} \left[
1 - \Gamma \left( \vec{b} - \vec{b} _{j} \right) 
\theta \left( {z} _{j}  - {z} \right) 
\right] 
\nonumber \\
& = & \exp i \chi \left( \vec{r} \left( \vec{b},z \right) , \vec{p}_{1} \right) \; .
\label{eq:profilefunc}
\end{eqnarray}
For the profile functions $\Gamma _{pN}$ for proton-nucleon scattering we
adopt the standard Gaussian parametrization 
\begin{displaymath}
\Gamma _{pN} (b) = \frac
{ \sigma ^{tot} _{pN} \left( 1  - i \epsilon _{pN} \right) }
{ 4 \pi \beta _{pN} ^{2} } \exp 
- \frac { b ^{2} } { 2 \beta _{pN} ^{2} } \; ,
\end{displaymath}
where $ \epsilon
_{pN}, \beta _{pN}, \sigma ^{tot} _{pN} $ have been determined from
the database of proton-proton and proton-neutron cross sections
\cite{Ryckebusch:2003fc}.

\begin{figure}
\includegraphics[width=0.45\textwidth]{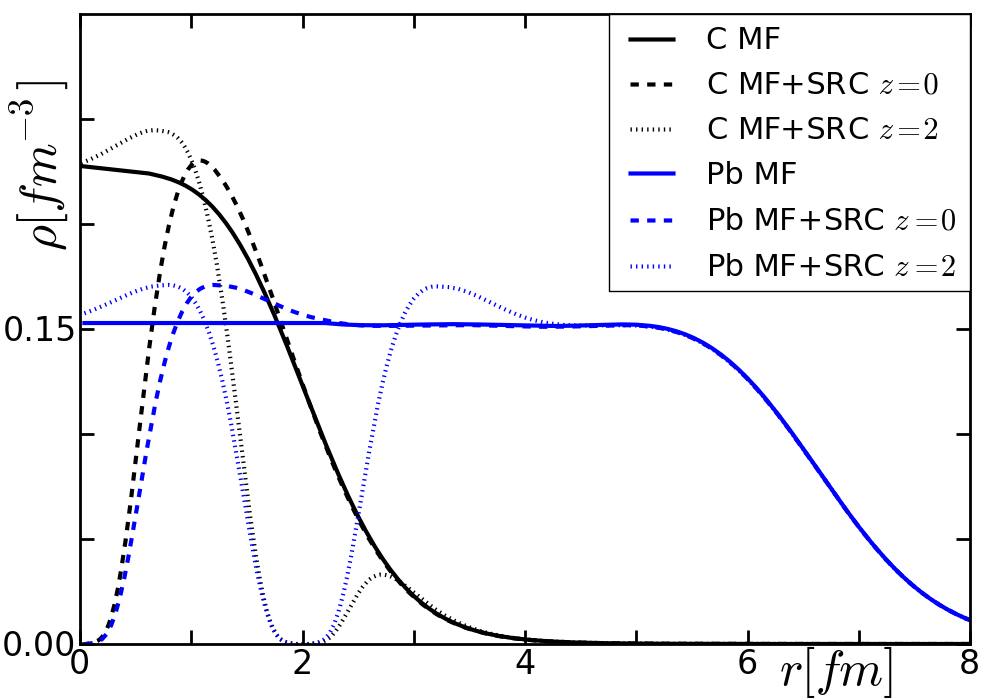}
\caption{(Color online) Comparison of the mean-field (MF) density
  $\rho _{A} ^{[1]} ( \vec{r} )$ and the SRC corrected effective
  density (denoted as MF+SRC) $\rho _{A} ^{eff} ( \vec{r}, 
  \vec{r} ' = (0, 0 ,z ) )$ for the target nuclei $^{12}$C and $
  ^{208}$Pb.  The dashed (dotted) lines are for $z=0$ ($z=2$).  
}
\label{fig:denscompare}
\end{figure}

It can be expected that the nucleon-nucleon interactions entering
Eq.~(\ref{eq:profilefunc}) by means of the profile function $\Gamma
_{pN} $ will be subject to medium modifications
\cite{Hillhouse:1998zz, ISI:000278399700004}.  Mechanisms like
Pauli blocking often lead to an effective reduction of the
nucleon-nucleon cross sections in the medium. At higher energies, the
effect of Pauli blocking is expected to become small. Another
important source of medium effects in the treatment of ISI and FSI,
are short-range correlations (SRC) 
\cite{Bianconi:1995mz, Alvioli:2008rw, PhysRevC.45.1863}. The SRC are related to
the finite size of the nucleons and the liquid properties of the
nucleus.  Indeed, the presence of a nucleon at some position $\vec{r}$
induces local fluctuations in the nuclear density.  These local
fluctuations, that go beyond the mean-field picture, can be included
in the numerical evaluation of the operators $
\widehat{\mathcal{S}}_{\text{RMSGA}}^ {(e,e'p)} (\vec{r})$ and $
\widehat{\mathcal{S}}_{\text{RMSGA}}^ {(p,2p)} (\vec{r})$, where
$\vec{r}$ is the interaction point of the impinging beam. With the
Eq.~(\ref{eq:profilefunc}), the $
\widehat{\mathcal{S}}_{\text{RMSGA}}^ {(p,2p)} (\vec{r}) $ of
Eq.~(\ref{eq:srmsgap2p}) has a very intuitive interpretation: it
represents the accumulated phase of one incoming and two outgoing
waves which are subject to a medium with $(A-1)$ grey disks
characterized by the profile function $\Gamma _{pN} $ and distributed
over the medium by a density distribution of the mean-field type $
\prod _{i=2} ^{i=A} \left| \phi _{\alpha _{i} } \left( \vec{r} _{i}
\right) \right| ^{2} $.  The latter expression for the density
distribution can be corrected for SRC, by using the information that
for a reaction to take place a nucleon should be present at the
initial interaction point \cite{Cosyn:2007er}.  This can be achieved
in the following way.  First, the squared single-particle wave
functions in the Eqs.~(\ref{eq:srmsgap2p}) and (\ref{eq:srmsgaeep})
can be approximated by the one-body density of the target nucleus
$\rho_A ^{[1]}(\vec{r})$ (normalized as $\int d \vec{r} \rho_A
^{[1]}(\vec{r}) = A $)
\begin{equation}
\mid \phi _ {\alpha _{i}} ( \vec{r}_{i} ) \mid ^{2}  \longrightarrow 
\frac { \rho _{A} ^{[1]}  ( \vec{r}_{i} ) } {A}  = 
\frac { \sum _{i=1} ^{i=A}  
\left| \phi _{\alpha _{i} } \left( \vec{r} _{i} \right) \right| ^{2} }
      {A} \; . 
\label{eq:srceq1}
\end{equation}
This substitution has a relatively minor impact on the computed effect
of ISI/FSI in the RMSGA model \cite{Ryckebusch:2003fc}.  Without any
loss of generality the $\rho_A ^{[1]}(\vec{r})$ can be replaced by the
ratio of the two-body density $\rho^{[2]}_A$ (normalized as $ \int
d\vec{r}_1 \int d \vec{r}_2 \rho^{[2]}_A(\vec{r}_1,\vec{r}_2)=A(A-1)$)
and the one-body density:
\begin{equation}\label{eq:subs}
  \rho^{[1]}_A(\vec{r}_2) \rightarrow
\frac{A}{A-1}\frac{\rho^{[2]}_A(\vec{r}_2,\vec{r})}{\rho^{[1]}_A(\vec{r})}\,,
\end{equation}
where $\vec{r}$ is the coordinate of the probe-target interaction.  For an
uncorrelated two-body density one has:
\begin{equation}
\rho^{[2]}_{A,\text{uncorr.}}(\vec{r}_1,\vec{r}_2)\equiv\frac{A-1}{A}\rho^{[1]}
_A(\vec { r } _1)\rho^{[1]}_A(\vec{r}_2)\,,
\end{equation}
 and Eq. (\ref{eq:subs}) becomes trivial.  One can implement the
 effect of central SRC in the two-body density by adopting the following
 functional form \cite{Cosyn:2007er,Frankel:1992er}
\begin{equation}
 \rho^{[2]}_{A,\text{corr.}}(\vec{r}_1,\vec{r}_2)
\equiv\frac{A-1}{A}\gamma(\vec{r}_1)\rho^{[1]}_A(\vec{r}_1)\rho^{[1]}_A(\vec{r}
_2)\gamma(\vec{r}_2) g(r_{12})\,,
\end{equation}
where $g(r_{12})$ is the so-called Jastrow correlation function 
\cite{Roth:2010bm}
and
$\gamma(\vec{r})$ a function which guarantees the proper normalization
of $\rho _{A} ^{[2]} $.  The $\gamma ( \vec{r} ) $ can be numerically
obtained as the solution of an integral equation.  With the above
expression for the two-body density, the Eq.  (\ref{eq:subs}) becomes
\begin{equation}
 \rho^{[1]}_A(\vec{r}_2) \rightarrow \gamma(\vec{r}_2) \rho^{[1]}_A(\vec{r}_2)
\gamma(\vec{r})
g(|\vec{r}_2-\vec{r}|) \equiv \rho^{\text{eff}}_A(\vec{r}_2, \vec{r})
\, .
\end{equation}
From the above derivations it follows that the computation of ISI/FSI
can be corrected for SRC by replacing $\left| \phi _ {\alpha_{i}}
\left( \vec{r}_{i} \right) \right|^2$ with
$\rho^{\text{eff}}_A(\vec{r}_{i}, \vec{r})/A$ in the
Eqs.~(\ref{eq:srmsgap2p}) and (\ref{eq:srmsgaeep}). The presence of a
nucleon at the interaction point $\vec{r}$ induces local fluctuations
in the probability distributions of the remaining $A-1$
nucleons. Within a radius $ r \le r_{N}$, where $r_{N}$ is the radius
of a nucleon, of the initial interaction point, there will be a
reduced probability to find a nucleon to scatter from, whereas for $r
\approx 2 r_{N}$ there will be an enhanced probability.  The sole
input required to determine $ \rho^{\text{eff}}_A(\vec{r}_2, \vec{r})
$ from $ \rho^{[1]}_A(\vec{r}_2) $ is the Jastrow correlation function
$g(r_{12})$. As this function is related to the short-range dynamics
of nuclei it is considered ``universal''. We use a $g(r_{12}) $ that
has a hard core extending over 0.8~fm and a second bump with a peak at
$r_{12} \approx 1.3$~fm. With this choice for the $g(r_{12}) $ we
obtained a fair description of the SRC contribution to the exclusive
$(e,e'pp)$ cross sections from $^{12}$C \cite{Blomqvist:1998gq} and
$^{16}$O \cite{ISI:000222562800013}. In Fig.~\ref{fig:denscompare} we
display for $^{12}$C and $^{208}$Pb the effective density for a
nucleon that is hit by an external probe in the center of the target
nucleus ($z$=0~fm) and in a more peripheral location along the
$z$-axis at 2~fm from the center ($z=2$~fm) of the target nucleus. The
SRC induce a hole in the density at the position of the probe-target
interaction point, and some enhancement at distances $\approx 1.5 $~fm
further away.  The proposed method for implementing the effect of SRC
in Glauber calculations leads to effective densities (Fig.~
\ref{fig:denscompare}) that are qualitatively very similar to those
produced in ab-initio calculations (see for example Fig.~1 of
Ref.~\cite{claudio2010}). Therefore, we consider the proposed method
to account for SRC in the ISI/FSI calculation, as realistic and
efficient.

\begin{figure}
\includegraphics[width=0.5\textwidth]{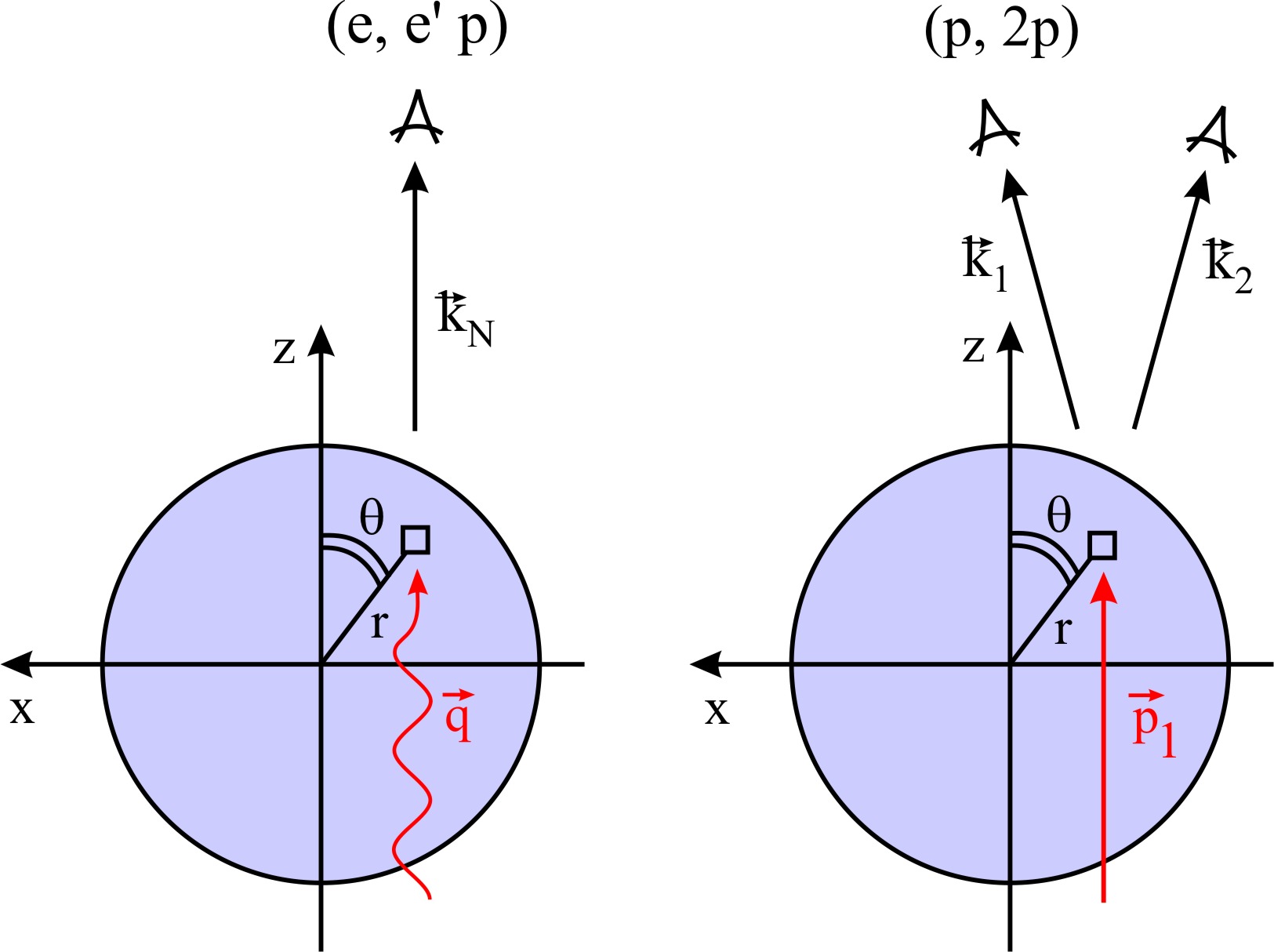}
\caption{Schematic representation of the $A(e,e'p)$ reaction in
  parallel kinematics and the $A(p,2p)$ reaction in coplanar and
  symmetric kinematics. The quantity $\delta (r , \theta) dr d \theta$
provides the contribution from the interval $ \left[ r + d r , \theta
  + d \theta \right] $ to the cross section. }
\label{fig:schemekin}
\end{figure}

\begin{figure*}
\includegraphics[width=15.6cm]{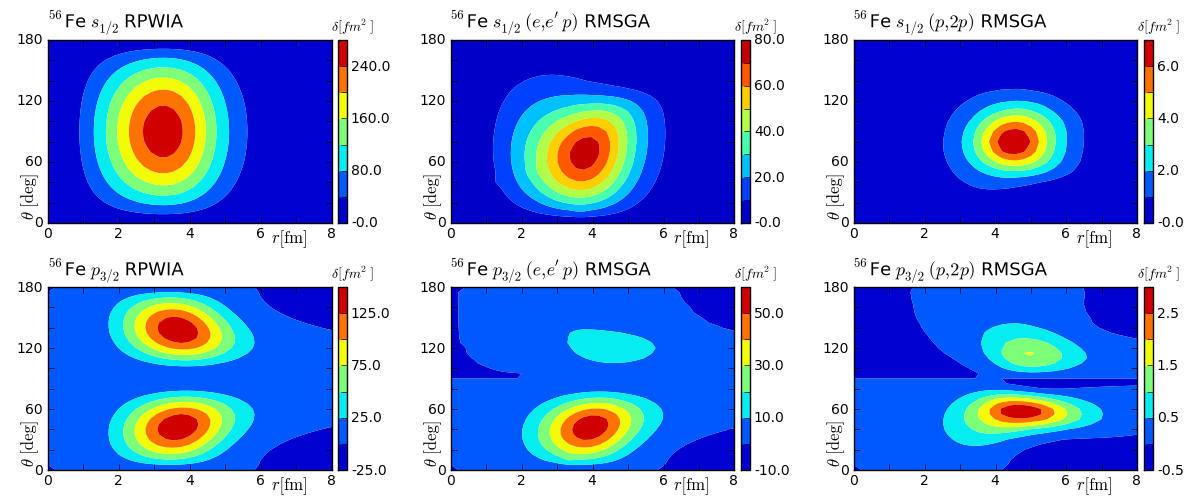}
\includegraphics[width=15.6cm]{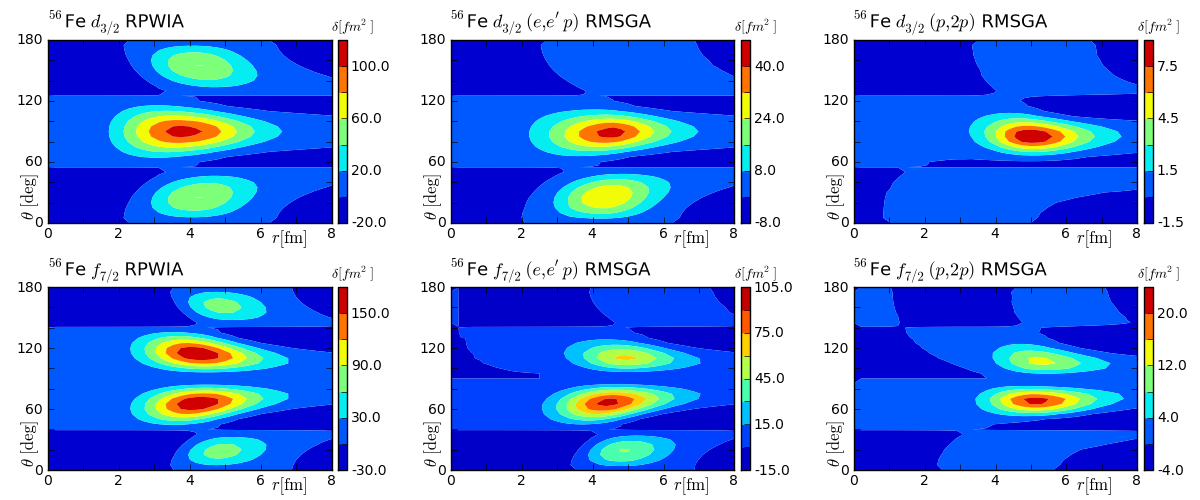}
\caption{{(Color online)} The function $\delta (r , \theta)$
  for knockout from various shells in the $^{56}$Fe target nucleus and
  ejected nucleon kinetic energies of 1.5~GeV. The $(e,e'p)$ results
  are for parallel kinematics. For $(p,2p)$ coplanar and symmetric
  kinematics is considered.  The magnitude of the momentum transfer
  $\left| \vec{q} \right| $ is adjusted so as to probe the maximum of
  the momentum distribution. This corresponds with $p_m =0$ MeV for
  the $1s_{1/2}$,
$p_m =105$ MeV for the $1p_{3/2}$,
$p_m =145$ MeV for the $1d_{3/2}$, and 
$p_m =180$ MeV for the $1f_{7/2}$. For the sake of reference, the
  measured proton root-mean-square radius in $^{56}$Fe is $r_{rms}
  \approx 3.75$~fm \cite{Peterson1970}.}
\label{fig:fe561}
\end{figure*}

\begin{figure}
\includegraphics[width=0.4\textwidth]{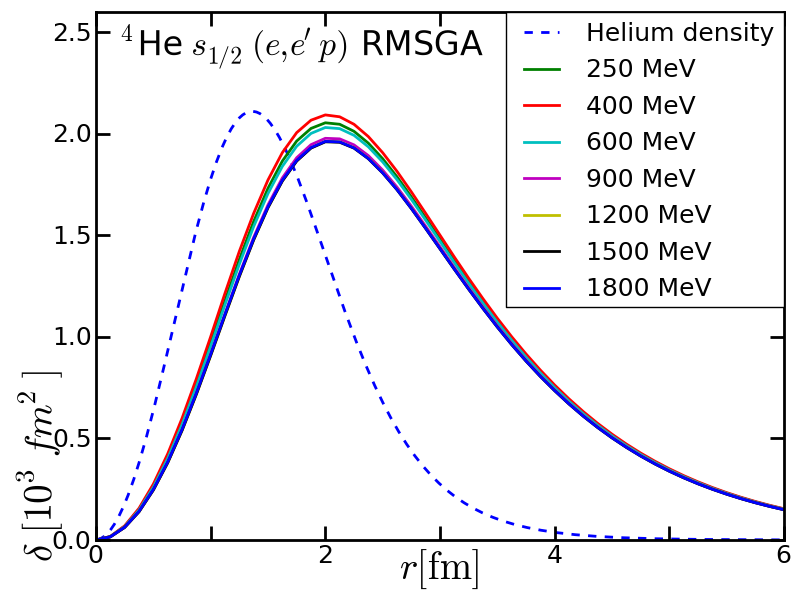}
\includegraphics[width=0.4\textwidth]{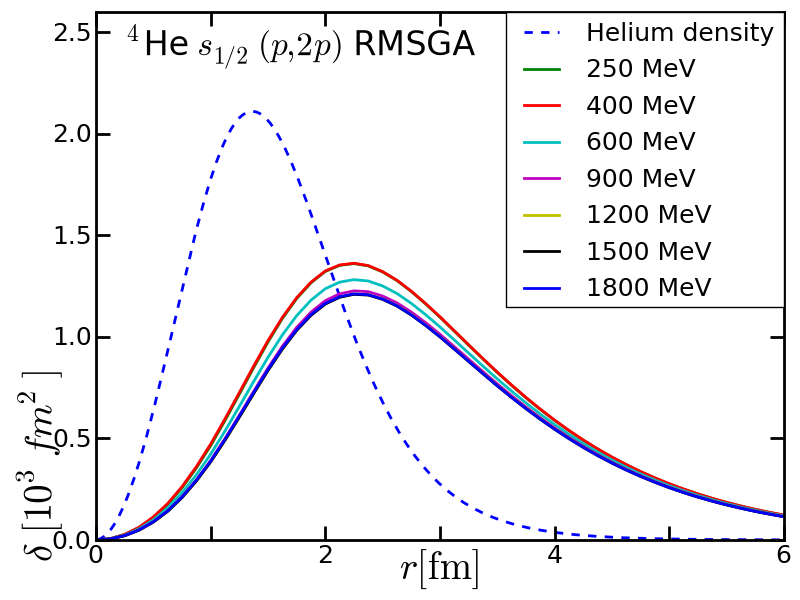}
\caption{(Color online) The energy dependence of the radial  
reaction probability densities $\delta (r) = \int d \theta
\delta ( r, \theta)  $. We consider proton knockout from
$^{4}$He at ejected proton kinetic energies of 250, 400, 600, 900, 1200, 1500,
and 1800~MeV. The $(e,e'p)$ ($(p,2p)$) RMSGA results are for parallel
(coplanar-symmetric) kinematics. For the sake of reference the 
$r^{2} \rho  ^{[1]}(r) $ for $^{4}$He is shown (not to scale!).}
\label{fig:edepen1}
\end{figure}

\begin{figure}
\includegraphics[width=0.5\textwidth]{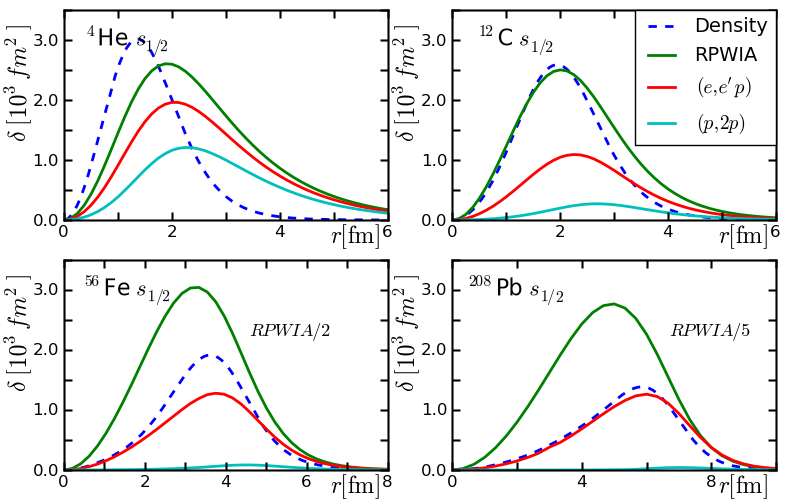}
\caption{(Color online) The target mass dependence of the
  radial reaction probability density $\delta (r) $. We consider proton
  knockout from the deep-lying $1s_{1/2}$ level and ejected proton
  kinetic energies of 1500~MeV. The magnitude of the momentum transfer
  $\left| \vec{q} \right| $ is adjusted so as to probe the maximum of
  the momentum distribution ($p_{m} = 0 $~MeV).
For the sake of reference we added
$r^{2} \rho _A (r) $ for the various target nuclei (not to scale!).
}
\label{fig:adepen1}
\end{figure}

\begin{figure}
\includegraphics[width=0.5\textwidth]{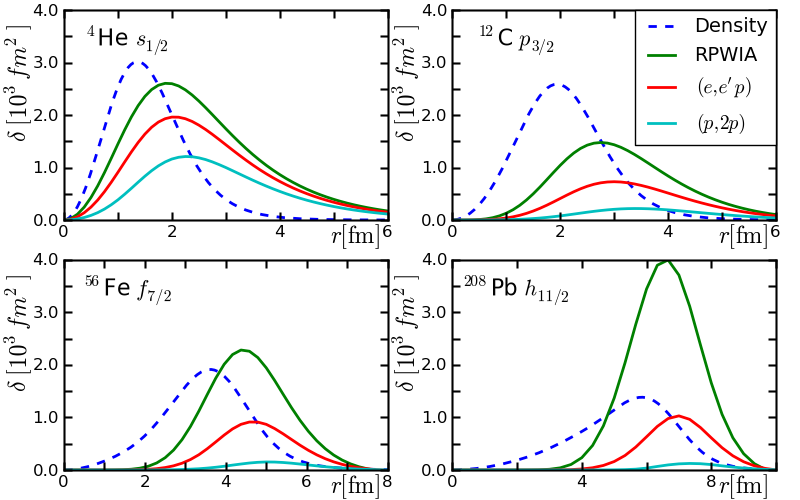}
\caption{(Color online) As in Fig.~\protect{\ref{fig:adepen1}} but now
    for knockout from one of the valence orbits.}
\label{fig:adepen2}
\end{figure}

\begin{figure*}
\includegraphics[viewport= 69 54 404 295,clip,width=0.33\textwidth]{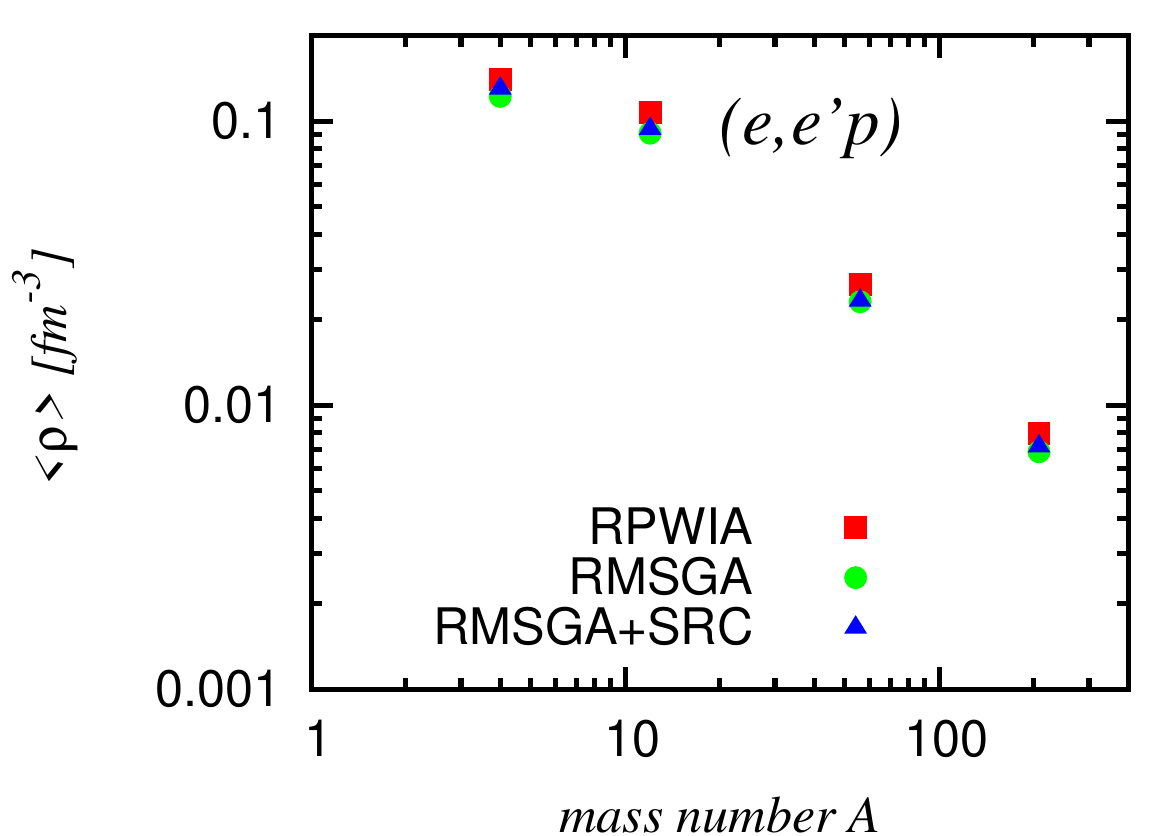}
\includegraphics[viewport= 69 54 404
  295,clip,width=0.33\textwidth]{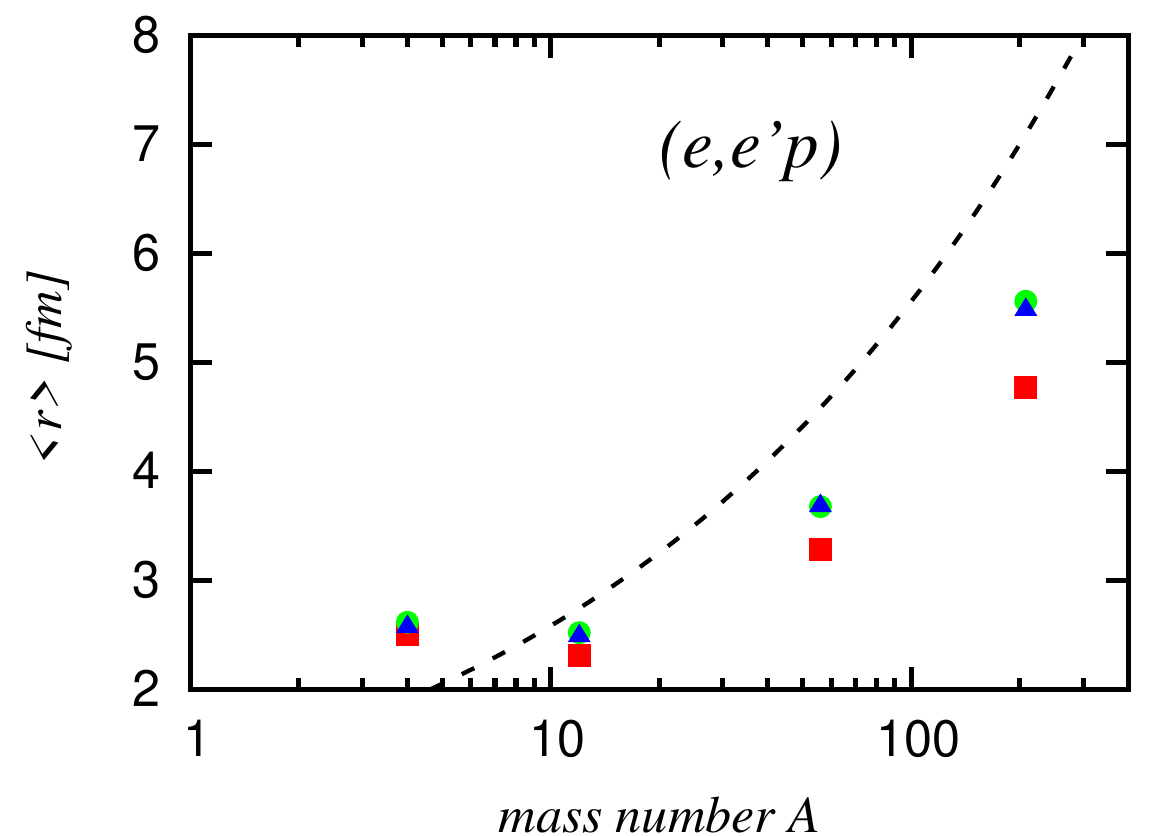} \\
\includegraphics[viewport= 69 54 404 295,clip,width=0.33\textwidth]{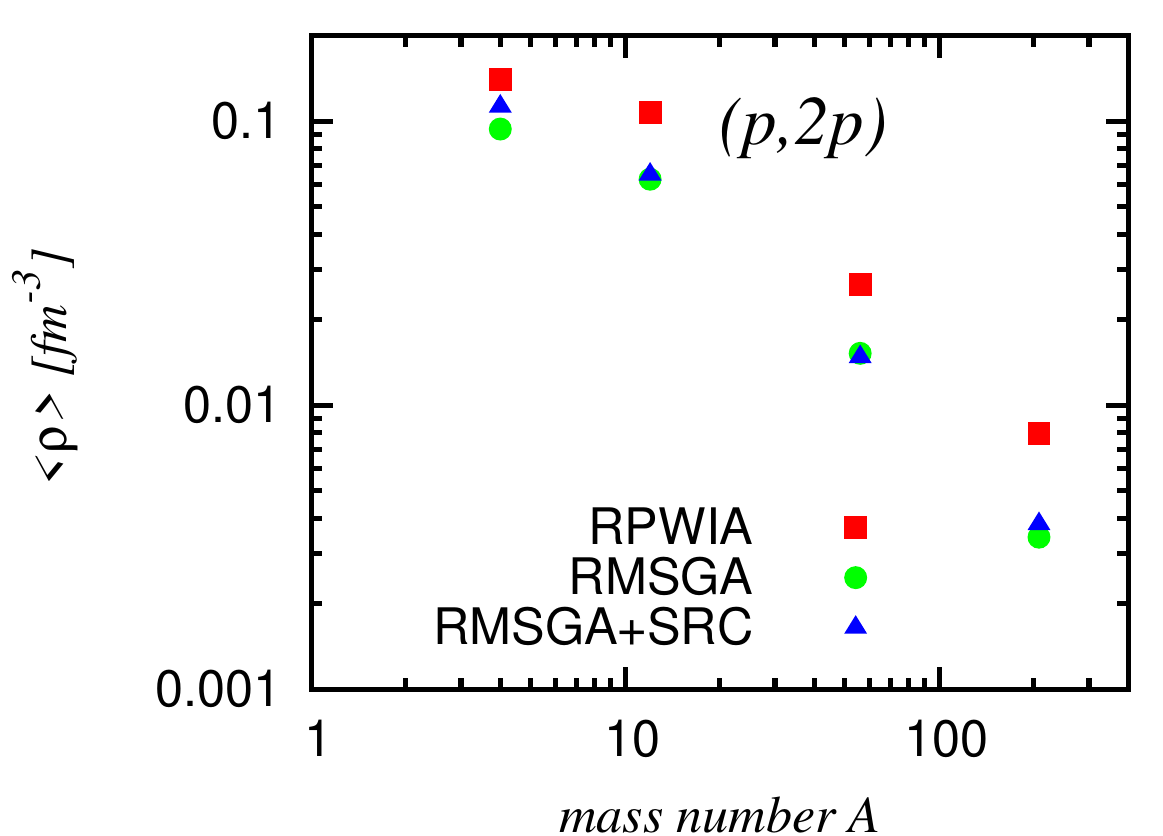}
\includegraphics[viewport= 69 54 404 295,clip,width=0.33\textwidth]{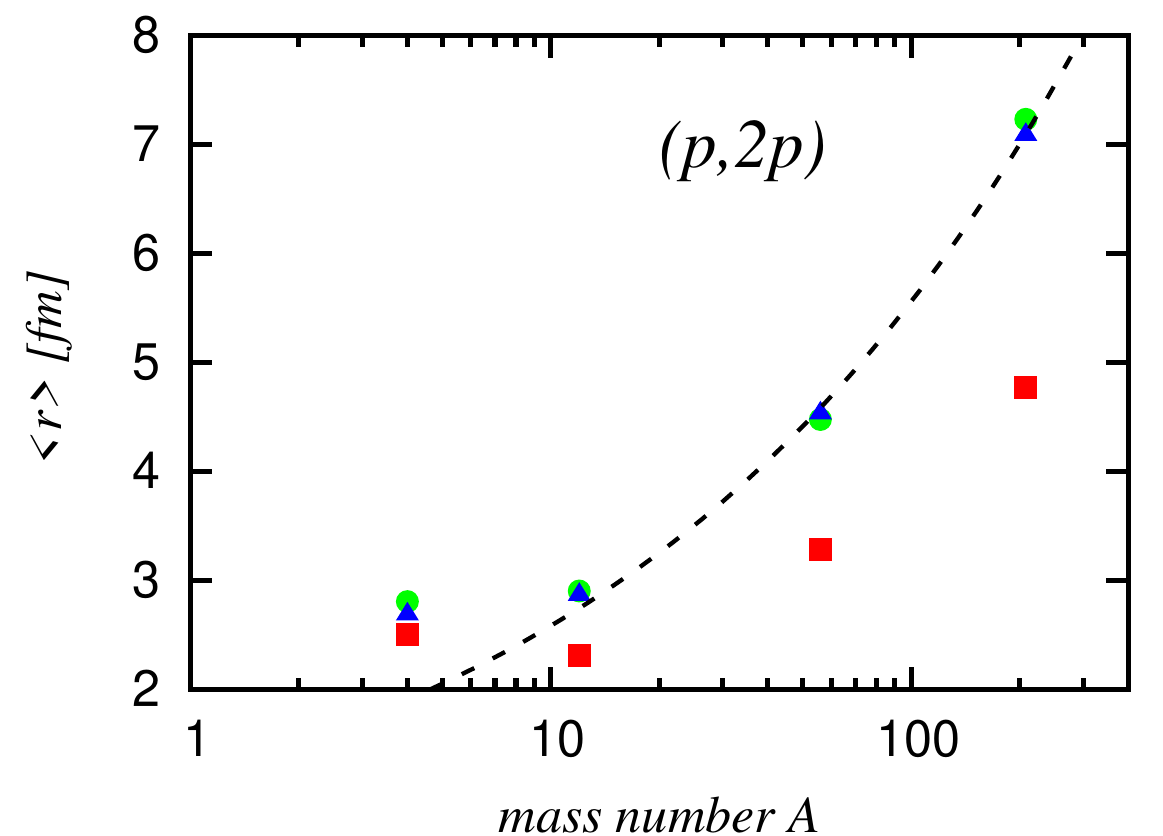}
\caption{(Color online) The target-mass dependence of the
  average density $\left< \rho \right> $ and the average radius
  $\left< r \right> $ that can be probed in single-nucleon knockout
  from the deep-lying $1s$ shell. The kinematic conditions are those
  from Fig.~\protect{\ref{fig:adepen1}}. The dotted line is $1.2 A ^{1/3}$.}
\label{fig:rdepen}
\end{figure*}

\begin{figure*}
\includegraphics[viewport= 69 54 404 295,clip,width=0.33\textwidth]{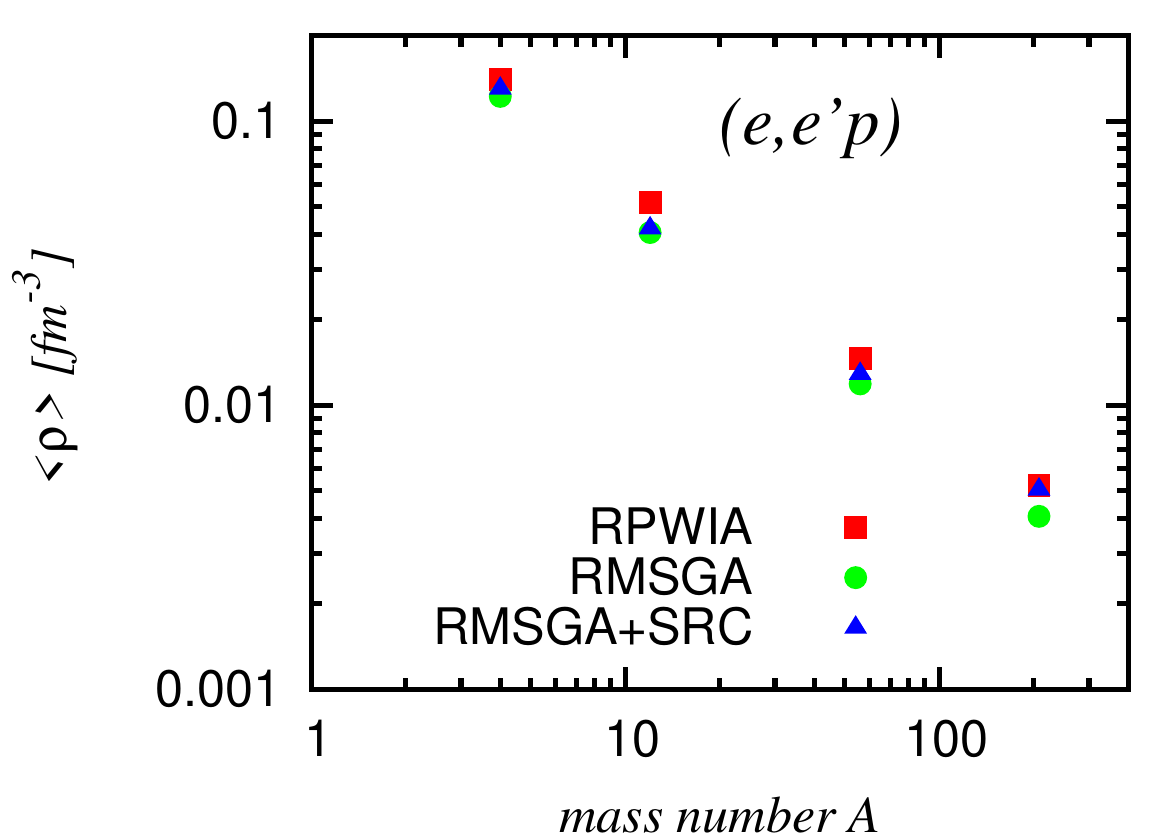}
\includegraphics[viewport= 69 54 404
  295,clip,width=0.33\textwidth]{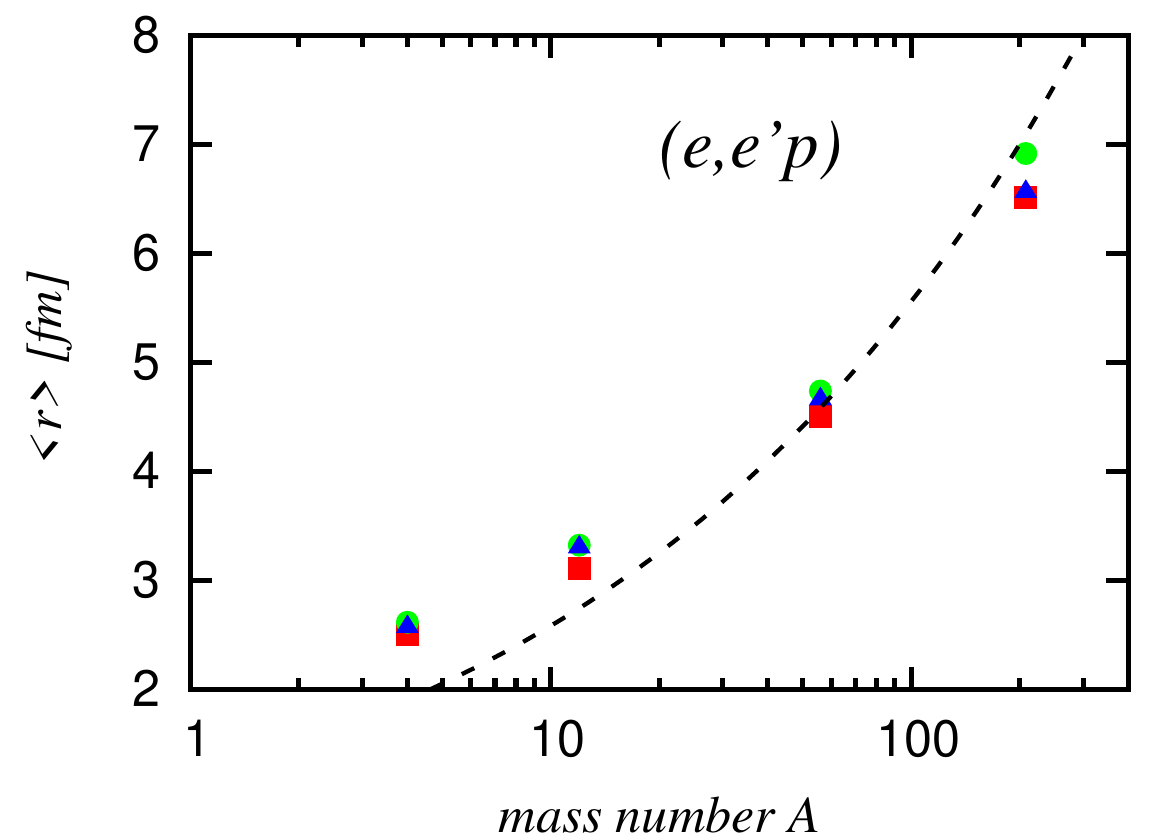} \\
\includegraphics[viewport= 69 54 404 295,clip,width=0.33\textwidth]{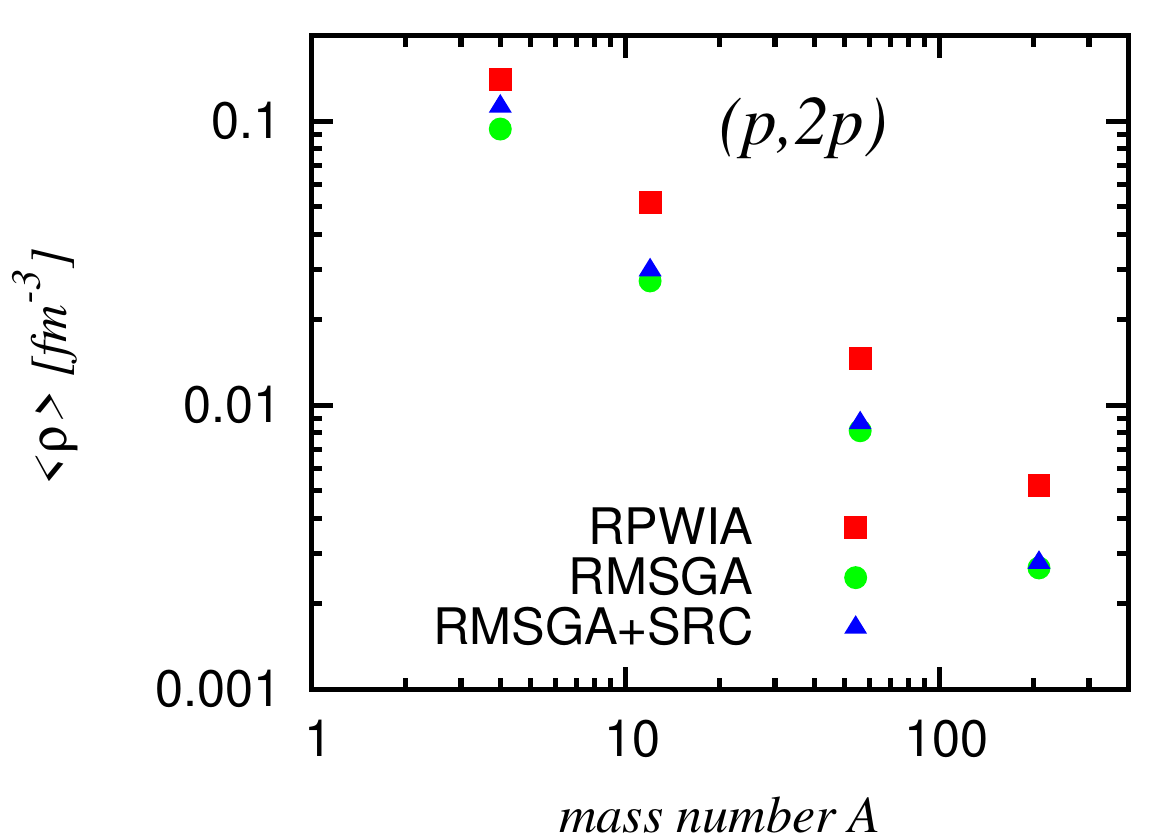}
\includegraphics[viewport= 69 54 404 295,clip,width=0.33\textwidth]{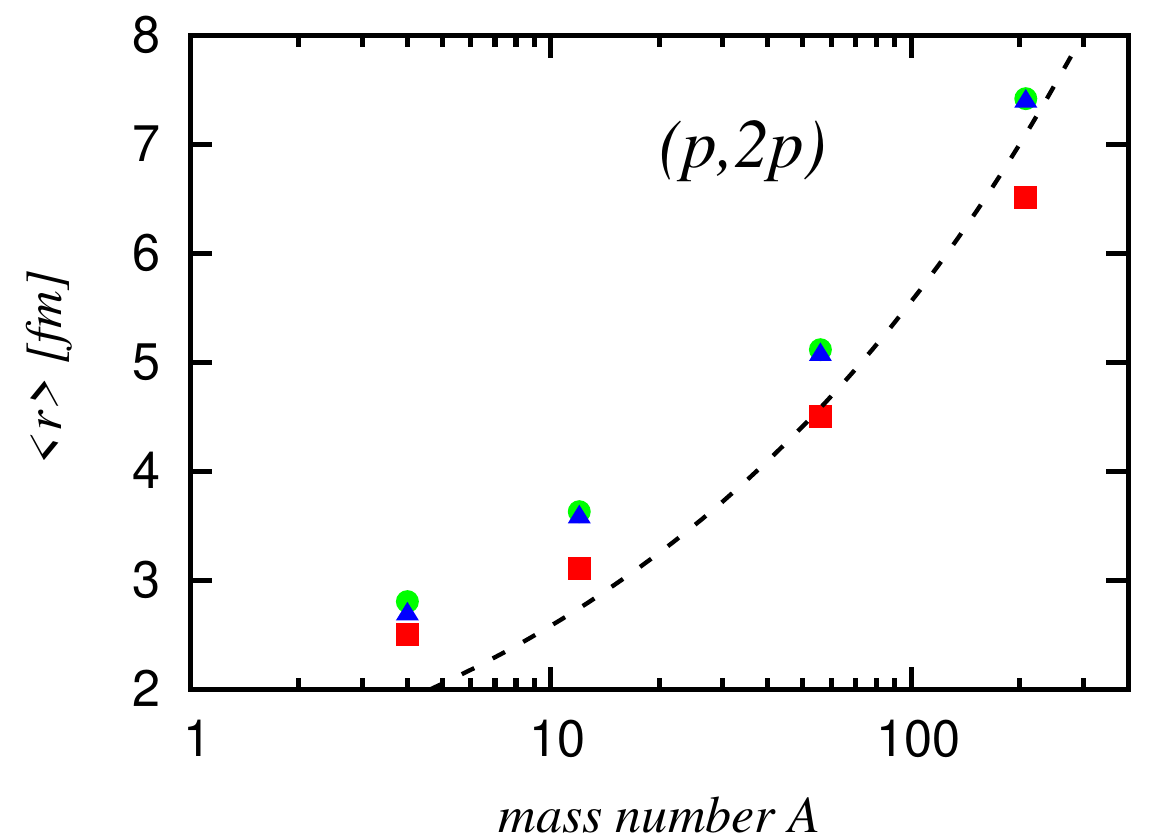}
\caption{(Color online) The target-mass dependence of the
  average density $\left< \rho \right> $ and the average radius
  $\left< r \right> $ which can be probed in single-nucleon knockout
  from the valence shell.  The kinematic conditions are those from
  Fig.~\protect{\ref{fig:adepen2}}. The dotted line is $1.2 A ^{1/3}$.  }
\label{fig:rhodepen}
\end{figure*}

\begin{figure}
\includegraphics[width=0.5\textwidth]{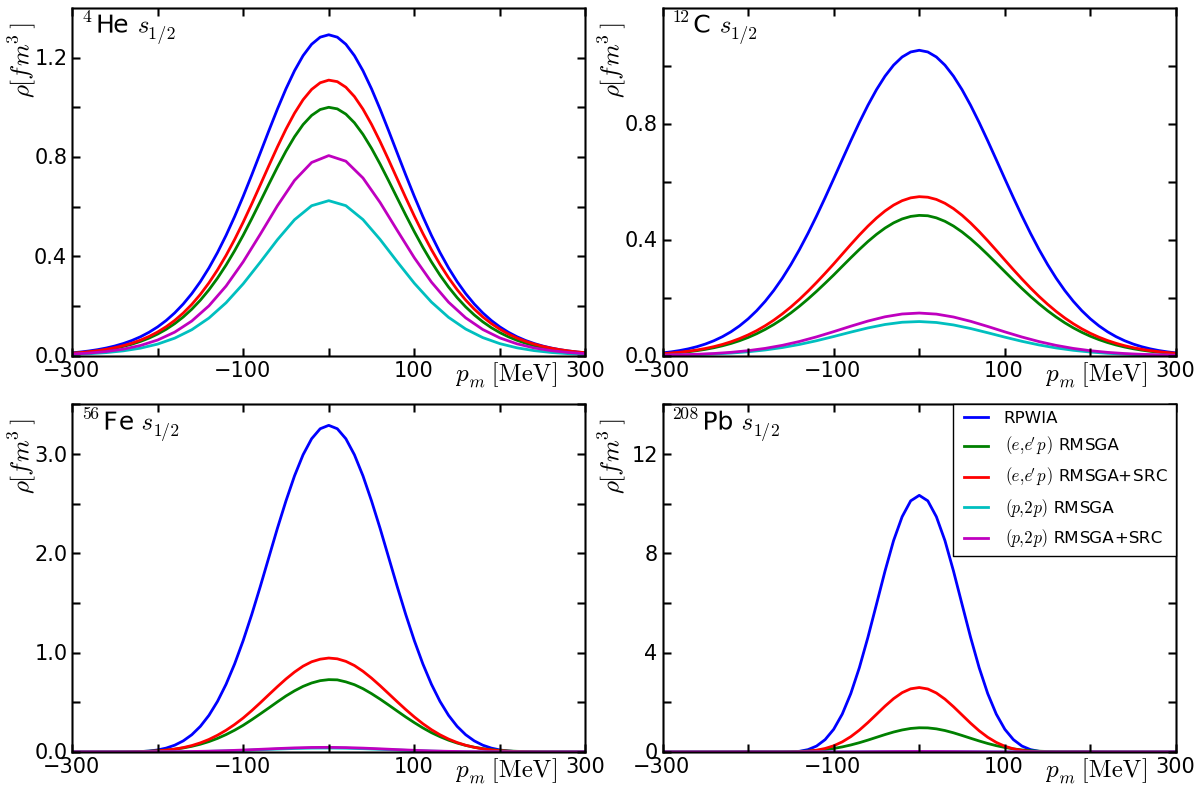}
\caption{(Color online) The RPWIA and RMSGA distorted momentum
    distribution for knockout from the deep-lying $1s_{1/2}$ orbit.  }
\label{fig:momentums}
\end{figure}

\begin{figure}
\includegraphics[width=0.5\textwidth]{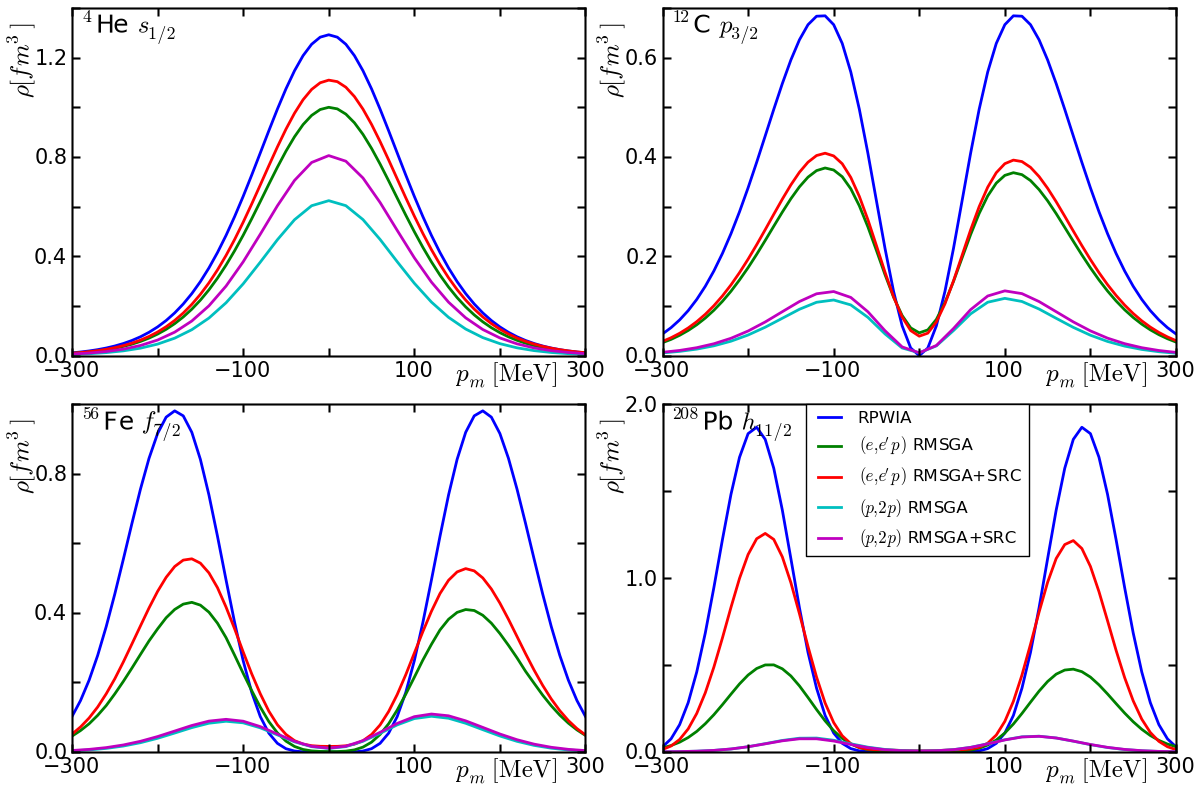}
\caption{(Color online) The RPWIA and RMSGA distorted momentum
    distribution for knockout from the valence shell.  }
\label{fig:momentumvalence}
\end{figure}

\begin{figure}
\includegraphics[viewport= 66 55 402 296, clip, width=0.4\textwidth]{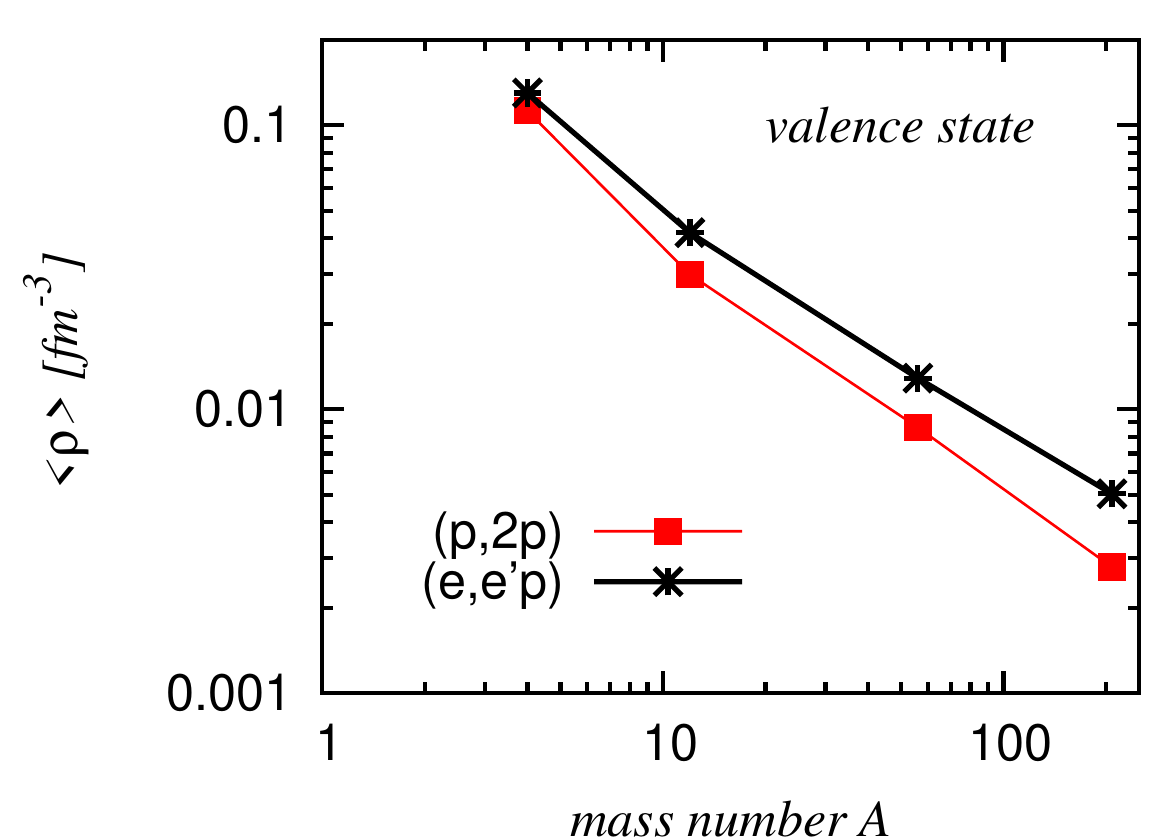}
\includegraphics[viewport= 66 55 402 296, clip, width=0.4\textwidth]{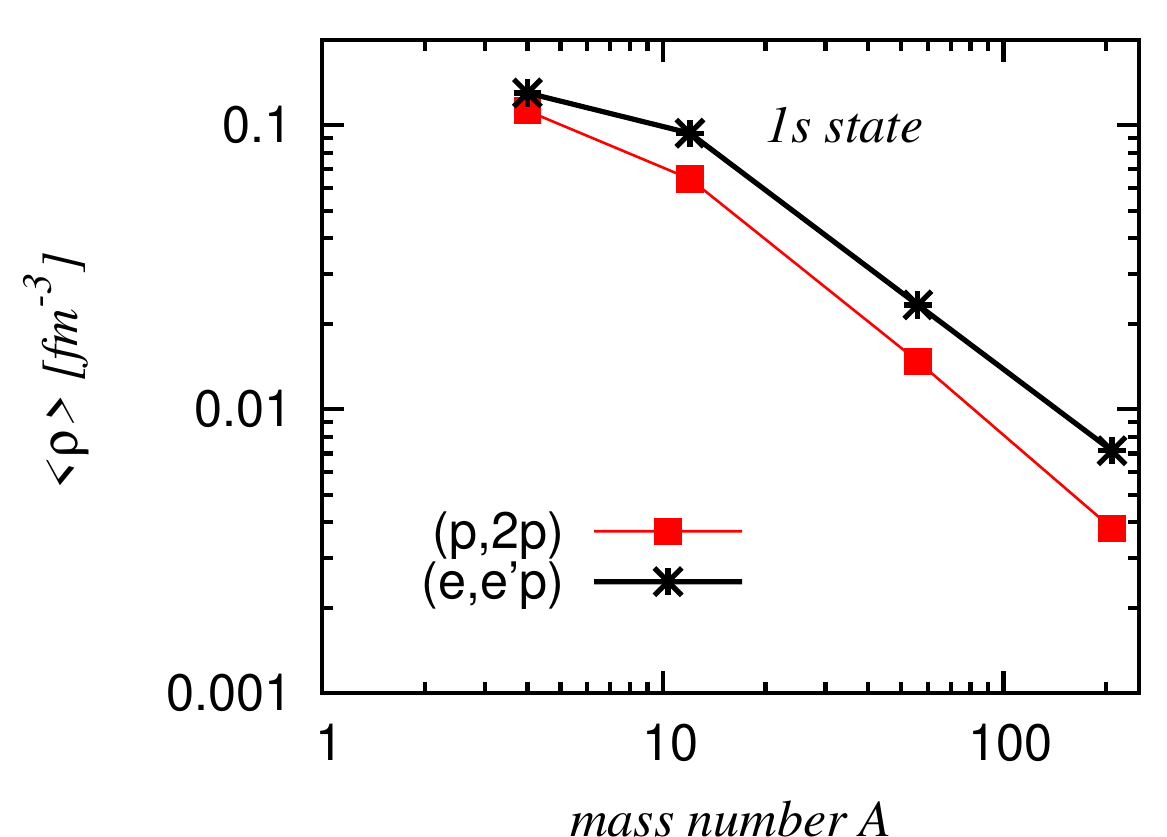}
\caption{The predicted average density which can be probed in
    a quasifree nucleon knockout reaction. The results are obtained in
the RMSGA framework and include the effect of SRC. }
\label{fig:final}
\end{figure}

\section{Results}
\label{sec:results}

We now present the results of the numerical calculations for $\delta
(r, \theta)$. The function $\delta ( r , \theta)$ depends on the
quantum numbers of the bound nucleon that collides with the proton or
electron beam and the kinematics of the reaction.  The considered
$A(e,e'p)$ and $A(p,2p)$ differential cross sections involve five kinematic
variables. The presented investigations aim at determining global
trends and choices with regard to the kinematics are in order.
For the $A(e,e'p)$ reaction we restrict ourselves to so-called
parallel kinematics: the final nucleon is detected along the direction
of the momentum transfer $\vec{q}$. For $A(p,2p)$ we consider coplanar
and symmetric kinematics: the two ejected nucleons have equal kinetic
energies ($\mid \vec{k}_1 \mid = \mid \vec{k}_2 \mid $) and escape
with equal opening angle but on opposite sides relative to the
direction of the momentum transfer. The considered kinematics is
illustrated in Fig.~\ref{fig:schemekin}. We stress that in many
respects there are strong analogies between the selected kinematics so
that meaningful comparisons between $(e,e'p)$ and $(p,2p)$ can be
made. 

The effect of ISI and FSI can be nullified by setting $
\widehat{\mathcal{S}}_{\text{RMSGA}} = 1 $ in
Eqs. (\ref{eq:srmsgap2p}) and (\ref{eq:srmsgaeep}). Under those
conditions one adopts the impulse approximation and all impinging and
ejected nucleons are described by plane waves. Accordingly, we will
refer to the corresponding reaction model as the relativistic
plane-wave impulse approximation (RPWIA). The difference between the
RPWIA and RMSGA predictions for $\delta (r , \theta) $ can be
exclusively attributed to nuclear attenuation.  In the calculations
including attenuation, we discriminate between the ``RMSGA'' and the
``RMSGA+SRC'' model variants. The latter includes the effect of SRC in
the modeling of the ISI/FSI.

In what follows we will display $(e,e'p)$ and $(p,2p)$ results for the
target nuclei $^{4}$He, $^{12}$C, $^{56}$Fe and $^{208}$Pb and use the
relativistic $\sigma \omega$ model to determine their single-particle
wave functions. 
We define the $z$-axis along the $\vec{q}$ and the
$xz$-plane as the reaction plane.  The ejected nucleons are detected
in the forward direction.    
We will start our discussions with the
results for the medium-heavy target nucleus $^{56}$Fe.

In Fig.~\ref{fig:fe561} we display the function $ \delta (r,\theta)$
for proton knockout from the $1s_{1/2}$, $1p_{3/2}$, $1d_{3/2}$, and
$1f_{7/2}$ orbits from $^{56}$Fe. We compare the $(p,2p)$ with the
$(e,e'p)$ result for an energy transfer of 1.5 GeV and conditions
probing the maximum of the undistorted momentum distribution 
\begin{equation}
\rho _{ n \kappa  }  (\vec{p}_{m}) = 
\sum_{s , m}
\left|  \int d \vec{r}
\frac { 
 e^{-i\vec{p} _{m} \cdot \vec{r}}
} 
{(2\pi)^{3/2}} 
\bar{u}(\vec{p}_m, s) 
\phi_{ n \kappa m } (\vec{r})
\right|^2 \; . 
\label{eq:freemomentumdistribution}
\end{equation}

The $x$-axis is a symmetry axis for $\delta (r, \theta )$ in the RPWIA
reaction picture.  Indeed, in the absence of nuclear attenuation, the
upper ($0 ^ {\circ} \le \theta \le 90^ {\circ} $) and lower hemisphere
($90 ^ {\circ} \le \theta \le 180^ {\circ} $) of the target nucleus
equally contribute to the measured signals in the detectors and the
$\delta (r,\theta)$ becomes equal for $(e,e'p)$ and $(p,2p)$. In
RPWIA the $\delta (r,\theta)$ reflects the symmetry imposed by the
quantum numbers $(n \kappa)$.  It is obvious that with increasing
orbital angular momentum $l$ the detected signals are increasingly
stemming from the peripheral areas of the target nucleus.  The ISI and
FSI have the strongest impact at the highest nuclear densities. This
reflects itself in the fact that the largest values of $\delta ( r ,
\theta)$ in the RMSGA model are shifted to larger values of $r$ in
comparison with what one finds in RPWIA.  As can be appreciated from
Fig.~\ref{fig:fe561} this shift is strongest for the deep-lying
$1s_{1/2}$ orbit and less pronounced for the valence $1f_{7/2}$
orbit. Further, attenuation breaks the symmetry between the
contribution from the upper and lower hemisphere as the first one is
positioned closer to the detector(s). The degree of asymmetry between
the contribution to the cross section from $(r , \theta)$ and $(r ,
180 ^{o} -\theta)$ is a measure for the impact of attenuation. Another
indicator is the ratio of the magnitude of RMSGA to the RPWIA
prediction for $\delta ( r, \theta) $.  Loosely speaking one could
interpret this ratio as the fraction of the available signal in the
target nucleus at some position $(r, \theta)$ that can withstand the
attenuating nuclear medium and makes it to the detectors.  Obviously,
the induced angular asymmetry, radial shift and overall reduction
occur for the $\delta (r,\theta) $ in both $(e,e'p)$ and $(p,2p)$.  All
three effects, however, are far more pronounced for the $(p,2p)$ than
for the corresponding $\delta (r, \theta)$ in $(e,e'p)$. Further, one
observes that the combined effect of ISI/FSI gradually diminishes as
one moves from the deep-lying to the valence single-particle orbits.

The preceding discussion concerned one particular proton kinetic
energy. Next, we report on the study of the energy dependence of the
impact of nuclear attenuation. As a representative example, in
Fig.~\ref{fig:edepen1} we show for $(e,e'p)$ and $(p,2p)$ the radial
reaction probability distribution $ \delta (r) = \int d \theta \delta
\left( r, \theta \right) $ for proton knockout from $^{4}$He at
kinetic energies ranging from 0.25 to 1.8~GeV. One observes a very
soft energy dependence in the radial dependence and the magnitude of
the $\delta (r) $. Therefore, we deem that the results for the average
densities and radii that will be presented below and that are obtained
at a specific kinetic energy of 1.5~GeV can be considered as
representative for nucleon knockout reactions with $T_{N} \gtrsim 250
$~MeV. In Figs.~ \ref{fig:edepen1}, \ref{fig:adepen1}, and
\ref{fig:adepen2} we have added the function $ r ^{2} \rho _{A} ^{[1]}( r)
$. The densities are computed with the basis of relativistic
single-particle wave functions which is also used in the reaction
dynamics calculations. The densities are plotted not to scale and help
in evaluating to what extent the knockout process succeeds in probing
the bulk regions of the target nucleus.

We wish to investigate the target-mass dependence of $\delta (r,
\theta)$. The results of Fig.~\ref{fig:fe561} pointed towards a strong
orbit dependence in the average radius and density that can be
probed. We consider knockout from the deep-lying $1s_{1/2}$
(Fig.\ref{fig:adepen1}) and one of the weakly-bound orbits
(Fig.\ref{fig:adepen2}) for the four representative target nuclei
considered in this work.  For $^{208}$Pb we opted for the $1h_{11/2}$
valence orbit as it is representative for orbits with large orbital
momentum. The $1h_{11/2}$ $\frac{11}{2} ^{-}$ hole state in $^{207}$Tl
is located at an excitation energy of 1.35~MeV.  Results of the RPWIA
and RMSGA calculations are contained in the Figs.~\ref{fig:adepen1}
and \ref{fig:adepen2}. We compare the $\delta (r) $ for the $(p,2p)$
reaction with the $(e,e'p)$ one.  The RPWIA result for $ \delta (r)$
is the reference figure of merit for knockout from a particular orbit.
The mass dependence can be best appreciated from the $1s_{1/2}$
knockout results from Fig.~\ref{fig:adepen1}.  The most spectacular
observation is the enormous decrease in the absolute value of the
RMSGA $(p,2p)$ radial reaction transition density for medium-heavy and
heavy nuclei.  For the $1s_{1/2}$ orbit, attenuation sheds about 90\%
of the $(p,2p)$ RPWIA signal in $^{12}$C, for $^{208}$Pb this becomes
more than 99\%. It is clear that for any meaningful extraction of
spectroscopic factors from $(p,2p)$ data the availability of a highly
reliable reaction model is of the utmost importance.  For knockout
from the valence orbits (Fig.~\ref{fig:adepen2}) the major fraction of
the signal stems from the tail of the density distribution of the
target nucleus.  For the light nuclei, the $(e,e'p)$ reaction performs
only slightly better than $(p,2p)$ when it comes to probing the bulk
regions. For the heavy nuclei, the growing role of attenuation makes
the reaction probability density increasingly peripheral, and it is
obvious that this primarily affects the $(p,2p)$.

In order to quantify the average density and radius that 
can be probed in the quasifree single-nucleon knockout reactions, 
we introduce the quantities \cite{PhysRevC.69.024604, PhysRevLett.78.1014}
\begin{eqnarray}
\left< \rho \right> & = & \frac 
{\int d r d \theta  \rho _{A} ^{[1]} 
\left( \vec{r} \right) \delta \left( r, \theta \right) }
{\int d r d \theta   \delta \left( r, \theta \right) } \; ,
\label{eq:averdens}
\\
\left< r \right> & = & \frac 
{\int d r d \theta  r   \delta \left( r, \theta \right) }
{\int d r d \theta   \delta \left( r, \theta \right) } \; ,
\label{eq:averradius}
\end{eqnarray}
where $\rho _{A} ^{[1]} \left( \vec{r} \right)$ is the density of the
target nucleus.  From the foregoing discussions we infer that for a
given target nucleus $A$ and orbit, the energy dependence of the
effective densities $ \left< \rho \right> $ and effective radius
$\left< r \right> $ is rather smooth.  For that reason we stick with
$T_{p}$=1.5~GeV and compile in Figs.~\ref{fig:rdepen} and
\ref{fig:rhodepen} the $\left< \rho \right> $ and $\left< r \right> $
for the $(e,e 'p) $ and $(p,2p) $ results contained in
Figs. \ref{fig:adepen1} and \ref{fig:adepen2}.  The RPWIA prediction
for the average radius $\left< r \right>$ increases with $A$ at a
slightly softer rate than $ A ^{1/3}$. Note that even in the idealized
attenuation-free world described by RWPIA, there is a strong mass
dependence in the average density that can be probed in a quasifree
nucleon knockout process. This strong mass dependence was exploited in
Ref.~\cite{PhysRevLett.78.1014} in order to probe the medium
dependence of the $NN$ scattering amplitude. In comparing extracted
information from $(e,e'p) $ and $(p,2p)$ reactions, like spectroscopic
factors, through the mass table, it is often not realized that the
reactions probe increasingly the target's surface region with growing
$A$. For proton knockout from $ ^{4}$He, RPWIA predicts an average
density $ \left< \rho \right> = 0.85 \rho _0 $, with the nuclear
saturation density $\rho _0 = 0.16$~fm$^{-3}$.  In $^{208}$Pb and
knockout from the $1h_{11/2}$ valence shell we find $ \left< \rho
\right> = 0.03 \rho _0$. All this before correcting for FSI/ISI which
further reduces the $ \left< \rho \right>$.

For light nuclei the RPWIA and RMSGA prediction for $\left<r \right>$ are
close. For heavier nuclei a different story emerges. For knockout from
the $1s_{1/2} $ orbit in $^{208}$Pb for example, attenuation makes the
$\left< r \right> $ to grow quite dramatically. Indeed, For $(p,2p)$ ($(e,e'p)$)
the RMSGA prediction is $\sim$ 2.5~fm ( $\sim$ 0.9~fm) larger than the
RPWIA value of 4.71~fm. For the valence states
(Fig.~\ref{fig:rhodepen}) the increase in $ \left< r \right> $ is not
larger than 0.4~fm in $(e,e'p)$.  In $(p,2p)$ the gain in $ \left< r
\right> $ is larger, but even in Pb it is smaller than 1 fm.  As
mentioned before, the $ \left< \rho \right> $ is decreasing
monotonically from the lightest to the heavier nuclei. Attenuation
makes that in reality smaller $ \left< \rho \right>$ will be
probed. The effect varies from a loss of couple of percent to a loss
in $ \left< \rho \right>$ of more than 50\%.  In $^{12}$C, the
$(p,2p)$ reaction from the $s1/2$ orbit can effectively probe higher
densities ($\left< \rho \right> =0.39 \rho _0$) than the $(e,e'p)$
reaction from the valence $p_{3/2}$ orbit ($\left< \rho \right> =0.25
\rho _0$).  For the $^{12}$C$(p,2p)$ reaction with knockout from the
$s1/2$ orbit the RMSGA prediction of $ \left< \rho \right>= 0.39 \rho
_0$ is comparable to the DWIA result $ \left<
\rho \right>= 0.34 \rho _0$ of Ref.~\cite{PhysRevC.69.024604} which
is obtained for 1 GeV incoming protons.  This illustrates the
robustness of the results of this work.

The results for $\left< \rho \right>$ and $\left< r \right> $ in
Figs.~\ref{fig:rdepen} and \ref{fig:rhodepen} allow one to estimate
the role of SRC in the modeling of the attenuation. In line with the
observations of Refs.~\cite{Bianconi:1995mz} and \cite{Frankel:1992er}
the SRC make the nucleus somewhat more transparent.  This reflects
itself in the fact that after including the SRC the RMSGA predictions
for $\left< \rho \right> $ and $ \left< r \right> $ are shifting
towards the RPWIA values. We wish to emphasize that the results of
Figs. \ref{fig:rdepen} and \ref{fig:rhodepen} refer to kinematics
corresponding with the maximum of the single-particle momentum
distributions.  On the basis of the densities shown in
Fig.~\ref{fig:denscompare} one may be tempted to expect rather
spectacular effects from SRC in the ISI/FSI.  The effect of SRC
on the $ \left< \rho \right>$ and $ \left< r \right> $ is rather
moderate due to the fact that the ISI/FSI are long-ranged in the
longitudinal direction and short-ranged in the transverse direction
\cite{Bianconi:1995mz}.

Up to this point we have evaluated quasifree processes for which the
kinematics is tuned to probe the maximum of the momentum
distribution. Now, we turn to a study of the distorted momentum
distribution $ \rho _{n\kappa} ^{D} ( p_{m} ) $ as a function of the
missing momentum $p_{m}$. We stick with parallel kinematics for
$(e,e'p)$ and coplanar and symmetric kinematics in $(p,2p)$. The
variation in $p_{m}$ is achieved by varying the kinetic energies of
the ejectiles at a fixed value of the momentum transfer $\vec{q}$. We
study knockout from the deep-lying $1s_{1/2}$ level for the various
nuclei in order to get some feeling about the mass dependence of the
attenuation.  As one can appreciate from Fig.~\ref{fig:momentums} with
growing $A$ the RPWIA and RMSGA predictions for $ \rho _{n\kappa} (
p_{m} ) $ increasingly diverge. A similar remark applies to the $ \rho
_{n\kappa} ( p_{m} ) $ for $(e,e 'p)$ and $ (p,2p)$.  For the valence
states (Fig.~\ref{fig:momentumvalence}) one observes similar trends,
though less pronounced. This can be easily understood by considering
that they probe the peripheral areas of the target nucleus.

At large missing momenta $ p _{m}$ the effect of the SRC corrections
on the distorted momentum distributions can be extremely large
\cite{Bianconi:1995mz}. Here, we concentrate on low missing momenta 
and observe that SRC make the nucleus more transparent for the
emission of nucleons. This observations complies with the conclusions
of Refs.~\cite{Bianconi:1995mz} and \cite{Frankel:1992er}. The SRC do
not dramatically affect the $p_{m}$ dependence of the distorted
momentum distributions at low $p_{m}$ but have an effect on its
magnitude. This makes it of the utmost importance to use a correlated
Glauber approach in order to extract precise information about the
spectroscopic factors $S_{n \kappa}$.

\section{Conclusions}
\label{sec:conclusions}

In summary, we have used a relativistic framework to make a
comparative and consistent study of the effective nuclear densities
which can be probed in quasifree 
$(p,2p)$ and $(e,e'p)$ reactions at high energies. We use relativistic
single-particle wave functions from the $ \sigma \omega$ model and a
relativistic extension of Glauber multiple-scattering theory.
Adopting the impulse and factorization factorization, both the
$(e,e'p)$ and the $(p,2p)$ cross sections are proportional to the
distorted momentum distribution $\rho _{n \kappa} ^{D}$ which include
the effect of nuclear attenuation for the impinging and ejected
nucleons.  The $\rho _{n \kappa} ^{D}$ reflects the effective momentum
density for a bound nucleon with quantum numbers $n \kappa$ that is
accessible for a certain reaction. We use this quantity as a tool to
make a quantitative assessment of the role of nuclear attenuation for
both types of reactions.  Obviously, nuclear attenuation is a complicating
factor in the extraction of nuclear-structure information from the
measured single-nucleon knockout signals. Strong attenuation has
a severe geometric influence as it makes the detected signal to carry
little information about the nuclear interior. An interesting
question, therefore, is how efficient  $(p,2p)$ reactions are in
probing the nuclear interior in comparison with the time-honored
$(e,e'p)$.

We have presented numerical results for knockout from one of the
weakly-bound levels and knockout from the deep-lying $1s$ level from
$^{4}$He, $^{12}$C, $^{56}$Fe, and $^{208}$Pb.  It emerges that to a
remarkable extent the effect of the nuclear attenuation on the angular
cross sections is independent of the energy of the initial and final
protons.  Accordingly, we have focused on one energy which we consider
representative.  The results for the average densities $\left< \rho
\right>$ which can be probed in quasifree single-nucleon knockout are
collected and shown in Fig.~\ref{fig:final}.  The $\left< \rho
\right>$'s are very sensitive to the quantum numbers of the bound
proton which collides with the probe. Further, there is strong
target-mass dependence. Whereas in the lightest nuclei one can probe
densities comparable to nuclear saturation density $\rho _ {0}$, for a
mid-heavy nucleus like Fe this is of the order of 10\% of $\rho _{0}$
and even smaller average densities are probed for a heavy nucleus like
Pb. For light nuclei, the $(p,2p)$ and $(e,e'p)$ reactions are
comparably efficient in probing the nucleus' inner regions. In
$^{208}$Pb, the effect of attenuation is very dramatic for $(p,2p)$ 
and the average density which can actually be probed can be half of
the $(e,e'p)$ one.

Our model for the ISI and FSI implements the effect of short-range
correlations. In line with previous studies we find that the SRC tend
to reduce the influence of attenuation.  The SRC moderately affect the
average density and radius that can be probed. We do find, however,
that after correcting for SRC the cross sections can
become substantially larger. The SRC corrections are particularly
relevant for the heavier target nuclei and the channels which probe
the inner-bound nucleons. As spectroscopic factors are typically
obtained from the ratio of the measured and the computed distorted
momentum distribution,  the SRC should become an essential ingredient
of any model for nuclear attenuation.

This work was supported by the Fund for Scientific
Research Flanders.

\bibliography{QFSextended}

\end{document}